\documentclass[aps, prd,twocolumn,superscriptaddress,nofootinbib,preprintnumbers]{revtex4-1}
\def\aj{AJ}%
\def\apj{ApJ}%
\def\apjs{ApJS}%
\def\aap{A\&A}%
\def\mnras{MNRAS}%
\def\prd{Phys.~Rev.~D}%
\def\nat{Nature}%
\def\physrep{Phys.~Rep.}%
\def\procspie{Proc.~SPIE}%
\newcommand{\om}{\Omega_\mr m}
\newcommand{\omb}{\Omega_\mr b}
\newcommand{\ns}{n_s}
\newcommand{\w}{w_0}

\def\galk{w^{\delta_{\rm g} \kappa_{\rm CMB}}(\theta) }

\def\sheartk{w^{\gamma_{\rm t} \kappa_{\rm CMB}}(\theta)}

\def\kk{w^{\kappa_{\rm CMB}\kappa_{\rm CMB}}(\theta)}

\def\galgal{w^{\delta_{\rm g} \delta_{\rm g}}(\theta)}
\def\galshear{w^{\delta_{\rm g} \gamma}(\theta)}

\def\shearshear{w^{\gamma \gamma}(\theta)}
\def\wtheta{w^{\delta_{\rm g} \delta_{\rm g}}(\theta)}
\def\gglensing{w^{\delta_{\rm g} \gamma}(\theta)}

\def\3x2pt{3$\times$2pt}
\def\5x2pt{5$\times$2pt}
\def\6x2pt{6$\times$2pt}

\newcommand{\planck}{\textit{Planck}}

\newcommand{\kcmb}{\kappa_{\rm CMB}}

\newcommand{\gammat}{\gamma_{\rm t}}

\newcommand{\delg}{\delta_{\rm g}}

\newcommand{\redmagic}{\textsc{redMaGiC}}
\newcommand{\redmapper}{\textsc{redMaPPer}}

\usepackage{amsmath,amssymb,latexsym,times}
\usepackage{numdef}
\usepackage{slashed}
\usepackage{graphicx}
\usepackage{grffile}
\usepackage[normalem]{ulem}
\usepackage{dcolumn}
\usepackage{xcolor}
\usepackage{booktabs}
\usepackage{amssymb}
\usepackage{slashed}
\usepackage{todonotes}
\usepackage{hyperref}
 \hypersetup{
     colorlinks=true,
     linkcolor=blue,
     citecolor = blue,      
     urlcolor=blue,
     }
\usepackage[T1]{fontenc}
\usepackage{ae,aecompl}
\usepackage{lineno}
\usepackage{eso-pic}% http://ctan.org/pkg/eso-pic
\usepackage{txfonts}
\newcommand{\be}{\begin{equation}}
\newcommand{\ee}{\end{equation}}
\newcommand{\ba}{\begin{eqnarray}}
\newcommand{\ea}{\end{eqnarray}}

\newcommand{\nside}{\ifmmode N_{\mathrm{side}}\else $N_{\mathrm{side}}$\fi}
\newcommand{\npix}{\ifmmode n_{\mathrm{pix}}\else $n_{\mathrm{pix}}$\fi}
\newcommand{\Npix}{\ifmmode N_{\mathrm{pix}}\else $n_{\mathrm{pix}}$\fi}
\newcommand{\lmin}{\ifmmode \ell_{\mathrm{min}}\else $\ell_{\mathrm{min}}$\fi}
\newcommand{\lmax}{\ifmmode \ell_{\mathrm{max}}\else $\ell_{\mathrm{max}}$\fi}

\newcommand{\Planck}{{\slshape Planck~}}

\newcommand{\mr}[1]{\mathrm{#1}}

\defcitealias{5x2methods}{B18}
\defcitealias{NKpaper}{O18a}
\defcitealias{GKpaper}{O18b}
\defcitealias{Krause:2017}{K17}
\defcitealias{Melchior:2017}{M17}
\defcitealias{Omori:2017}{O17}
\defcitealias{DESy1:2017}{DES-Y1-3x2}

\begin{document}
\title[\5x2pt Key Paper]
{Dark Energy Survey Year 1 Results: Joint Analysis of Galaxy Clustering, Galaxy Lensing, and CMB Lensing Two-point Functions}

\author{T.~M.~C.~Abbott}
\affiliation{Cerro Tololo Inter-American Observatory, National Optical Astronomy Observatory, Casilla 603, La Serena, Chile}
\author{F.~B.~Abdalla}
\affiliation{Department of Physics \& Astronomy, University College London, Gower Street, London, WC1E 6BT, UK}
\affiliation{Department of Physics and Electronics, Rhodes University, PO Box 94, Grahamstown, 6140, South Africa}
\author{A.~Alarcon}
\affiliation{Institut d'Estudis Espacials de Catalunya (IEEC), 08193 Barcelona, Spain}
\affiliation{Institute of Space Sciences (ICE, CSIC),  Campus UAB, Carrer de Can Magrans, s/n,  08193 Barcelona, Spain}
\author{S.~Allam}
\affiliation{Fermi National Accelerator Laboratory, P. O. Box 500, Batavia, IL 60510, USA}
\author{J.~Annis}
\affiliation{Fermi National Accelerator Laboratory, P. O. Box 500, Batavia, IL 60510, USA}
\author{S.~Avila}
\affiliation{Institute of Cosmology \& Gravitation, University of Portsmouth, Portsmouth, PO1 3FX, UK}
\author{K.~Aylor}
\affiliation{Department of Physics, University of California, Davis, CA, USA 95616}
\author{M.~Banerji}
\affiliation{Institute of Astronomy, University of Cambridge, Madingley Road, Cambridge CB3 0HA, UK}
\affiliation{Kavli Institute for Cosmology, University of Cambridge, Madingley Road, Cambridge CB3 0HA, UK}
\author{N.~Banik}
\affiliation{Fermi National Accelerator Laboratory, P. O. Box 500, Batavia, IL 60510, USA}
\author{E.~J.~Baxter}
\affiliation{Department of Physics and Astronomy, University of Pennsylvania, Philadelphia, PA 19104, USA}
% ...

\collaboration{DES \& SPT Collaborations}

\date{Last updated \today}

\label{firstpage}


\begin{abstract}
\newpage We perform a joint analysis of the auto and cross-correlations between three cosmic fields: the galaxy density field, the galaxy weak lensing shear field, and the cosmic microwave background (CMB) weak lensing convergence field. These three fields are measured using roughly 1300 sq. deg. of overlapping optical imaging data from first year observations of the Dark Energy Survey and millimeter-wave observations of the CMB from both the South Pole Telescope Sunyaev-Zel'dovich survey and \emph{Planck}. We present cosmological constraints from the joint analysis of the two-point correlation functions between galaxy density and galaxy shear with CMB lensing. We test for consistency between these measurements and the DES-only two-point function measurements, finding no evidence for inconsistency in the context of flat $\Lambda$CDM cosmological models. Performing a joint analysis of five of the possible correlation functions between these fields (excluding only the CMB lensing autospectrum) yields $S_{8}\equiv \sigma_8\sqrt{\Omega_{\rm m}/0.3} = 0.782^{+0.019}_{-0.025}$ and $\Omega_{\rm m}=0.260^{+0.029}_{-0.019}$. We test for consistency between these five correlation function measurements and the \emph{Planck}-only measurement of the CMB lensing autospectrum, again finding no evidence for inconsistency in the context of flat $\Lambda$CDM models. Combining constraints from all six two-point functions yields $S_{8}=0.776^{+0.014}_{-0.021}$ and $\Omega_{\rm m}= 0.271^{+0.022}_{-0.016}$. These results provide a powerful test and confirmation of the results from the first year DES joint-probes analysis. 
\end{abstract}

\preprint{DES-2017-0295}
\preprint{FERMILAB-PUB-18-512-AE}

\maketitle

%\date{Last updated \today}
%\pubyear{2017}

\section{Introduction}

The recent advent of wide-field imaging surveys of large-scale structure (LSS) enables observations of a rich variety of signals that probe dark matter, dark energy, the nature of gravity and inflation, and other aspects of the cosmology and physics of the Universe. Originally, most of the constraining power from LSS observations came from measurements of the luminous tracers of the underlying mass \citep[e.g.][]{Davis:1985, White:1993}. First detected in the year 2000 \citep{Bacon:2000,Kaiser:2000,VanWaerbeke:2000,Wittman:2000}, 
 galaxy weak lensing now provides valuable complementary information.  Weak lensing enables an almost direct measurement of the mass distribution, greatly enhancing the constraining power of LSS imaging surveys \citep[e.g.][]{Kilbinger:2013,Hildenbrandt:2017, DESy1:2017}. 
 
Although several galaxy imaging surveys have demonstrated  successful measurements of galaxy clustering and lensing, significant challenges with these measurements remain.  Particularly challenging are the inference of gravitationally induced galaxy shears (see \cite{Mandelbaum:2017} for a review) and the inference of galaxy redshifts from photometric data (see \cite{Salvato:2018} for a review).  Systematic errors in these measurements can lead to biased cosmological constraints.  As an illustration of these challenges, the recent cross-survey analysis of \cite{Chang:2018} showed that the cosmological constraints from several recent weak lensing surveys are in tension, and that result from these surveys can be significantly impacted by differences in analysis choices. Distinguishing possible hints of new physics from systematic errors is perhaps the main challenge of present day observational cosmology.

The Dark Energy Survey (DES) has adopted several strategies for combating sources of systematic errors and ensuring robustness of cosmological constraints to analysis choices. Two aspects of this approach are worth emphasizing here. First, DES has adopted a multi-probe approach, whereby multiple survey observables are analyzed jointly. By combining multiple probes in a single analysis, the results can be made more robust to systematic errors and analysis choices impacting any single observable. The cosmological analysis presented in \cite{DESy1:2017} (hereafter \citetalias{DESy1:2017}) considered a combination of three two-point functions formed between galaxy overdensity, $\delta_{\rm g}$, and weak lensing shear, $\gamma$. This combination --- which includes galaxy clustering, cosmic shear, 
and galaxy-galaxy lensing --- is particularly robust to possible systematics and nuisance parameters because each probe depends differently on expected systematics.  We refer to this combination of three two-point correlation functions as \3x2pt. A second aspect of the DES approach to ensuring robust cosmological constraints is adherence to a strict blinding policy, so that both measurements and cosmological constraints are blinded during analysis. Blinding becomes especially important when the impact of systematic errors and analysis choices becomes comparable to that of statistical uncertainty, as appears to be the case with some current galaxy surveys.

By extending the multi-probe approach to include correlations with lensing of the cosmic microwave background (CMB), the robustness and constraining power of cosmological constraints from LSS surveys can be further improved. Photons from the CMB are gravitationally deflected by the LSS, and the distinct pattern of the lensed CMB can be used to probe lensing structures along the line of sight (see \cite{Lewis2006} for a review). Experiments such as  the Atacama Cosmology Telescope (ACT; \cite{Swetz2011}), the \Planck satellite \citep{Tauber2010,Planck2011} and the South Pole Telescope (SPT; \cite{Carlstrom:2011}) make high resolution and low noise maps of the CMB, enabling measurement of the CMB lensing convergence, $\kappa_{\rm CMB}$ (ACT: \cite{Das:2011,vanEngelen:2015,Sherwin:2017}; {\it Planck}: \cite{Planck:cmblensing, Plancklensing:2018}; SPT: \cite{vanEngelen:2012, Story:2015, Omori:2017}).  With a combination of galaxy and CMB lensing, we effectively get to measure the lensing effects of LSS twice. Any difference between the lensing measurements performed with these two sources of light would likely be indicative of systematic errors.  Of course, CMB lensing is also impacted by sources of systematic error (see \cite{5x2methods} for a discussion in the context of the measurements presented here); however, the systematic errors impacting CMB lensing are very different from those faced by galaxy surveys.  For instance, measurement of CMB lensing is unaffected by source redshift uncertainty, shear calibration biases, and intrinsic alignments.  Since systematic errors will necessarily become more important as statistical uncertainties decrease,  joint analyses including CMB lensing are likely to be an important part of the analysis of data from future surveys, such as the Large Synoptic Survey Telescope \citep{LSSTsciencebook} and CMB Stage-4 \citep{Abazajian:2016}.

With these considerations in mind, we present here an extension of the \3x2pt analysis to include all correlations 
between $\delta_{\rm g}$, $\gamma$, and $\kappa_{\rm CMB}$, with $\kappa_{\rm CMB}$ measured by both the South Pole Telescope and the {\it Planck} satellite. We first perform a joint analysis of angular cross-correlations between $\delta_{\rm g}$ and $\kappa_{\rm CMB}$, and between $\gamma$ and $\kappa_{\rm CMB}$, which we refer to as $\galk$ and $\sheartk$, respectively. Using two statistical approaches, we test for consistency between this combination of probes and the \3x2pt data vector considered in the analysis of \citetalias{DESy1:2017}. Finding consistency, we perform a joint cosmological analysis of all five correlation functions, which we refer to as \5x2pt. We next test for consistency between the \5x2pt combination of probes and measurements of the autocorrelation of CMB lensing from \cite{Planck:cmblensing}.  Again finding consistency, we perform a joint cosmological analysis of all six two-point functions, which we refer to as \6x2pt.  Following \citetalias{DESy1:2017}, we adhere to a blinding policy whereby all significant measurement and analysis choices were frozen prior to unblinding.

This work represents the first complete cosmological analysis of two-point functions between DES observables and measurements of CMB lensing.\footnote{Refs.~\cite{kirk16}, \cite{Baxter:2016} and \cite{giannantonio16} also performed joint analyses of two-point functions between DES and CMB lensing, but these analyses only allowed at most one cosmological parameter to vary.} It uses first year observations from DES and CMB observations from both the 2008-2011 SPT Sunyaev-Zel'dovich survey (SPT-SZ) and {\it Planck}. We view this analysis as laying the foundations for future joint analyses of two-point functions between DES observables and CMB lensing.  Consequently, we have made several analysis choices (for both DES and SPT+{\it Planck} data) that ensure a high degree of robustness of the analysis, while sacrificing some statistical power. These choices are also well motivated given our focus on performing a consistency test of the \citetalias{DESy1:2017} results.  We comment in the Discussion section on a number of improvements we expect to implement with future datasets and analyses.  

\begin{figure*}
\begin{center}
\includegraphics[height=0.35\textwidth]
{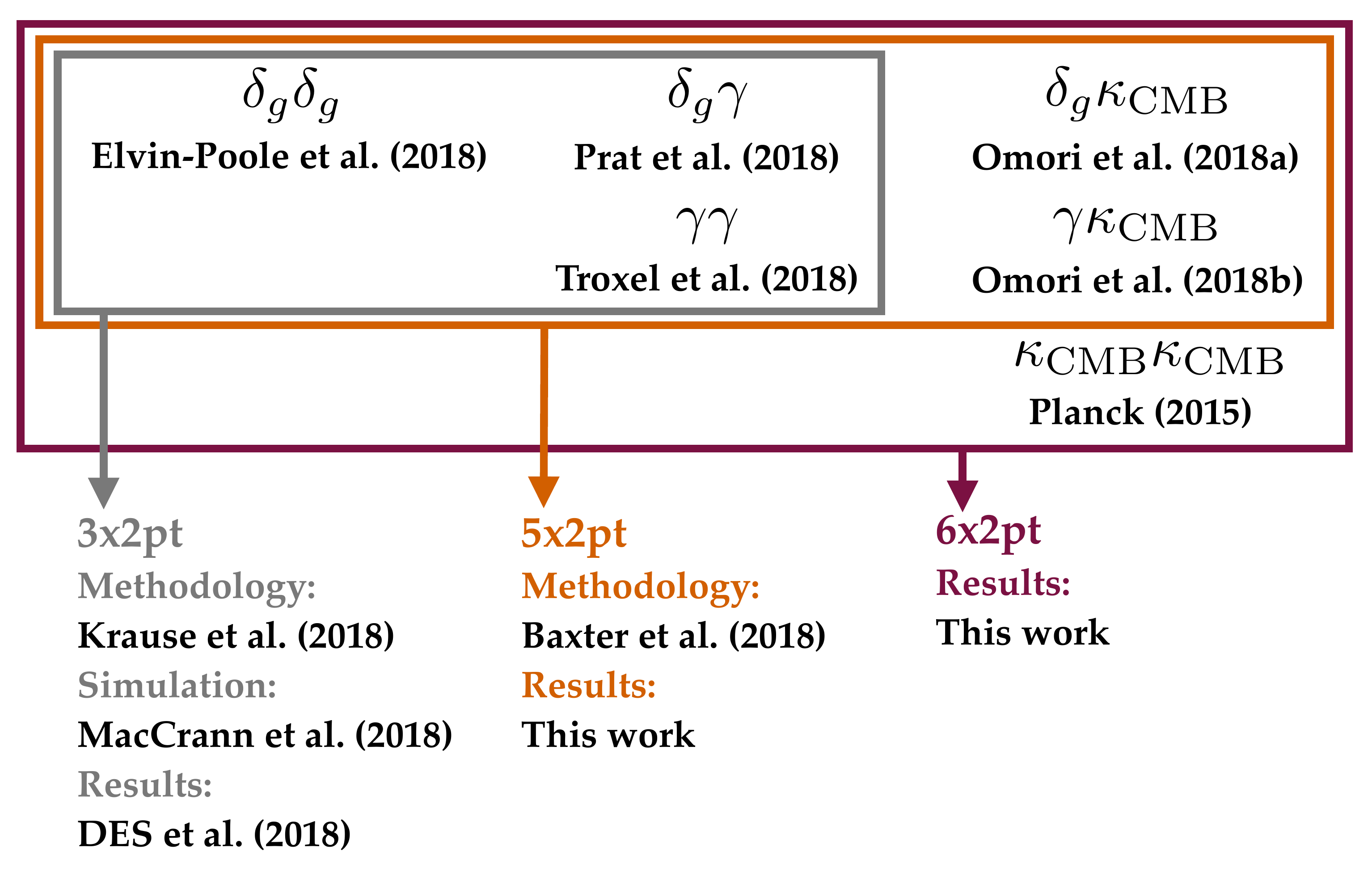}
\caption{Summary of papers presenting analyses of two-point functions of DES-Y1 measurements of projected galaxy density, $\delta_g$, and shear, $\gamma$, as well as cross-correlations with the CMB lensing maps, $\kappa_{\rm CMB}$, from \cite{Omori:2017}.  The blue box represents the joint \3x2pt analysis, while the orange and black boxes represent the \5x2pt and \6x2pt analyses considered in this work. }
\label{fig:paper_summary}
\end{center}
\end{figure*}

This work extends earlier analyses of cross-correlations between DES catalogs and SPT CMB lensing maps \citep{giannantonio16,Baxter:2016,kirk16}. Similar joint analyses of two-point functions between galaxies, galaxy shears, and CMB lensing have also been presented by \cite{Nicola2016} and \cite{Doux2017}.  The analysis presented here also relies heavily on several recent papers analyzing DES-Y1 data and cross-correlations with $\kappa_{\rm CMB}$ maps. The analysis of the \3x2pt combination of two-point functions presented in \citetalias{DESy1:2017} uses the two-point measurements from \cite{ElvinPoole2017, Prat2017, Troxel2017}, which in turn build on many ancillary measurement and methodological papers \citep{Cawthon2017,Davis:2017,Gatti:2017,Hoyle2017,Krause:2017,Zuntz17ca}.  Ref.~\cite{MacCrann:2018} applied the \3x2pt methodology to simulated datasets for the purposes of validation.  Ref.~\cite{5x2methods} (hereafter \citetalias{5x2methods}) extended the methodology of \cite{Krause:2017} (hereafter \citetalias{Krause:2017}) to include modeling of the cross-correlation between galaxies and shear with the $\kappa_{\rm CMB}$ map.  The details of the modeling of these two additional correlation functions, the characterization of potential systematics, and the motivation for angular scale cuts, are described in \citetalias{5x2methods}.  Ref.~\cite{NKpaper} (hereafter \citetalias{NKpaper}) present measurements of the correlation between galaxies and $\kappa_{\rm CMB}$, while
\cite{GKpaper} (hereafter \citetalias{GKpaper}) present measurements of the correlation between shear and $\kappa_{\rm CMB}$. The relations between the different two-point functions considered here and the relevant references are summarized in Fig.~\ref{fig:paper_summary}.

The structure of the paper is as follows: in Sec.~\ref{sec:model} we briefly summarize the model for the two-point function measurements; in Sec.~\ref{sec:data} we describe the data used in this analysis; in Sec.~\ref{sec:consistency_metrics} we give an overview of the metrics that we use to evaluate consistency between the different datasets considered in this work; in Sec.~\ref{sec:blinding} we describe our blinding scheme and validation tests; in Sec.~\ref{sec:results} we present the results of our cosmological analysis of the measured two-point functions; we conclude in Sec.~\ref{sec:discussion}.

\section{Model}
\label{sec:model}

\subsection{Correlation function model}
\label{sec:basic_model}

We model the set of \5x2pt correlation functions as described in \citetalias{Krause:2017} and \citetalias{5x2methods}. We present a brief overview of the modeling choices here and refer readers to those works for more details.  

We are interested in the two-point correlation functions between three fields: the projected galaxy overdensity, $\delta_{\rm g}$, the lensing shear measured from images of galaxies, $\gamma$, and the lensing convergence measured from the CMB, $\kappa_{\rm CMB}$.  To differentiate between the galaxies used as tracers of the matter density field (i.e. the samples used to measured $\delta_{\rm g}$) and the galaxies used as sources of light for measuring gravitational lensing (i.e. the samples used to measure $\gamma$), we will frequently refer to these samples as {\it tracers} and {\it sources}, respectively.  We use superscripts to indicate different redshift bin measurements of these fields.

Using the Limber approximation \citep{Limber:1953}, the harmonic-space cross-spectra between these three fields can be related to an integral along the line of sight over the matter power spectrum, with an appropriate weighting function. The use of the Limber approximation is justified given our choice of angular scales and redshift binning \citep{Krause:2017}.  We use $\kappa_{\rm s}$ to represent the lensing convergence defined from the source galaxy images to distinguish it from $\kappa_{\rm CMB}$.  It is convenient to first compute cross-spectra with the spin-0 $\kappa_{\rm s}$ field, and to subsequently convert into cross-correlations with components of the spin-2 shear field, $\gamma$.   We use the notation $f_{\alpha}$ to generically represent one of the fields $\delta_g$, $\kappa_{\rm CMB}$ and $\kappa_{\rm s}$.  The cross-spectra can then be written as \citep{Loverde:2008}
\begin{equation}
C^{f_{\alpha} f_{\beta}}(\ell) = \\\int d\chi \frac{q_{f_{\alpha}} (\chi) q_{f_{\beta}} (\chi) }{\chi^2} P_{\rm NL} \left( \frac{\ell+1/2}{\chi}, z(\chi)\right),
\label{eq:limber}
\end{equation}
where $\chi$ is the comoving distance and  $P_{\rm NL}(k,z)$ is the nonlinear matter power spectrum, which we compute using CAMB \citep{Lewis:2000}. In a spatially flat universe, the weight functions for the different fields are
\begin{equation}
q_{\kappa_{\rm s}^i} = \frac{3 \Omega_{\rm m} H_0^2}{2c^2}\frac{\chi}{a(\chi)}\int_{\chi}^{\chi_{\rm h}} d\chi' \frac{n^i_{\rm s}(z(\chi')) \frac{dz}{d\chi'}}{\bar{n}_{\rm s}^i} \frac{\chi' - \chi}{\chi'},
\end{equation}
\begin{equation}
\label{eq:qkcmb}
q_{\kappa_{\rm CMB}} = \frac{3\om H_0^2}{2c^2}\frac{\chi}{a(\chi)}  \frac{\chi^* - \chi}{\chi^*},
\end{equation}
\begin{equation}
\label{eq:deltag_weight}
q_{\delta_{\rm g}^i} = b_g^i \frac{n_{\rm g}^i(z(\chi))}{\bar{n}_{\rm g}^i} \frac{dz}{d\chi}, 
\end{equation}
where $n^i_{\rm s}(z)$ and $n^i_{\rm g}(z)$ are the redshift distributions of source and tracer galaxies in the $i$th bin, and $\bar{n}_{\rm s}^i$ and $\bar{n}_{\rm g}^i$ are the corresponding integrated number densities in this redshift bin.  In Eq.~\ref{eq:deltag_weight} we have assumed linear galaxy bias with a single bias parameter, $b_{\rm g}^i$, for each galaxy redshift bin $i$.  

The position-space correlation functions can be related to the harmonic-space cross-spectra as follows.  The correlations of the galaxy density field with itself and with the CMB convergence field are computed via
\begin{eqnarray}
w^{\delta_g^i \delta_g^j}(\theta) &=& \sum \frac{2\ell + 1}{4\pi} P_{\ell} (\cos(\theta)) C^{\delta_g^i \delta_g^j}(\ell)\\
w^{\delta_g^i \kappa_{\rm CMB}}(\theta) &=& \sum \frac{2\ell + 1}{4\pi} F(\ell) P_{\ell} (\cos(\theta)) C^{\delta_g^i \kappa_{\rm CMB}}(\ell), 
\end{eqnarray}
where $P_{\ell}$ is the $\ell$th order Legendre polynomial, and $F(\ell)$ describes filtering applied to the $\kappa_{\rm CMB}$ map. For correlations with the $\kappa_{\rm CMB}$ map of \cite{Omori:2017} (hereafter \citetalias{Omori:2017}), we set $F(\ell)= B(\ell) \Theta(\ell - 30) \Theta(3000 - \ell)$, where $\Theta(\ell)$ is a step function and $B(\ell) = \exp (-\ell(\ell + 1)/\ell_{\rm beam}^2)$ with $\ell_{\rm beam} \equiv \sqrt{16 \ln 2}/\theta_{\rm FWHM} \approx 2120$.  The motivation for this filtering is discussed in more detail in \citetalias{5x2methods}.

We compute the cosmic shear two-point functions, $\xi_{+}$ and $\xi_{-}$, using the flat-sky approximation:
\begin{equation}
\xi^{ij}_{+/-}(\theta) = \int \frac{d\ell \, \ell}{2\pi}  J_{0/4}(\ell \theta)  C^{\kappa_s^i \kappa_s^j}(\ell),
\end{equation}
where $J_i$ is the second order Bessel function of the $i$th kind.  For ease of notation, we will occasionally use $w^{\gamma\gamma}$ to generically refer to both $\xi_{+}$ and $\xi_{-}$.

When measuring the cross-correlations between galaxies and shear, or between $\kappa_{\rm CMB}$ and shear, we consider only the tangential component of the shear field, $\gamma_{\rm t}$.  These correlation functions are then given by
\begin{eqnarray}
w^{\delta_{\rm g}^i \gamma_{\rm t}^j }(\theta) &=& \int \frac{d\ell \, \ell}{2\pi} J_2(\ell \theta)  C^{\delta_{\rm g}^i \kappa_{\rm s}^j }(\ell), \\
w^{\gamma_{\rm t}^i \kappa_{\rm CMB}}(\theta) &=& \int \frac{d\ell \, \ell}{2\pi} F(\ell) J_2(\ell \theta)  C^{\kappa_s^i \kappa_{\rm CMB}}(\ell).\label{eq:shearkcorr}
\end{eqnarray}

In addition to the coherent distortion of galaxy shapes caused by gravitational lensing, galaxies can also be intrinsically aligned as a result of gravitational interactions.  We model intrinsic galaxy alignments using the nonlinear linear alignment (NLA) model \citep{bridle07}, which modifies $q_{\kappa_{\rm s}^{i}}$ as: {
\begin{equation}
q_{\rm \kappa_{\rm s}^i}(\chi) \rightarrow q_{\kappa_{\rm s}^i}(\chi) - A(z(\chi)) \frac{n_{\rm s}^i((z(\chi))}{\bar{n}_{\rm s}^i} \frac{dz}{d\chi},
\end{equation}
where
\begin{equation}
A(z) = A_{{\rm IA}, 0} \left( \frac{1+z}{1+z_0} \right)^{\eta_{\rm IA}} \frac{0.0139\Omega_{\rm m} }{D(z)},
\end{equation}
and where $D(z)$ is the linear growth factor and $z_0$ is the redshift pivot point which we set to  $0.62$ as done in \citetalias{Krause:2017}.
}

We also model two sources of potential systematic measurement uncertainties in our analysis: biases in the photometric redshift estimation, and biases in the calibration of the shear measurements. Photometric redshift bias is modeled with an additive shift parameter, $\Delta z$, such that the true redshift distribution is related to the observed distribution via $n_{\rm true}(z) = n_{\rm obs}(z-\Delta z)$. We adopt separate redshift bias parameters $\Delta z_{\rm g}^i$ and $\Delta z_{\rm s}^i$ for each tracer and source galaxy redshift bin, respectively. 

We model shear calibration bias via a multiplicative bias parameter, $m^i$, for the $i$th redshift bin.  We then make the replacements
\begin{eqnarray}
\xi^{ij}_{+/-}(\theta) &\rightarrow& (1+m^i)(1+m^j)\, \xi^{ij}_{+/-} (\theta)\\
w^{\gamma_{\rm t}^i \kappa_{\rm CMB}}(\theta) &\rightarrow& (1+m^i) w^{\gamma_{\rm t}^i \kappa_{\rm CMB}}(\theta).
\end{eqnarray}

\subsection{Choice of angular scales}
\label{sec:scale_cuts}

In our modeling of the correlation functions we neglect several physical effects, including nonlinear galaxy bias, the impact of baryons on the matter power spectrum, and contamination of the CMB maps by the thermal Sunyaev-Zel'dovich (tSZ) effect. As shown in \cite{Troxel2017}, \citetalias{Krause:2017} and \citetalias{5x2methods}, some of these effects can have a significant impact on the measured correlation functions at small scales.  

In order to reduce the impact of unmodeled effects on our analysis, we restrict analysis of the correlation function measurements to angular scales where their impact is small. The choice of angular scale cuts for the \3x2pt data vector was motivated in \cite{Troxel2017} and \citetalias{Krause:2017}, while the choice of angular scale cuts for $\galk$ and $\sheartk$ was motivated in \citetalias{5x2methods}.  In the case of the latter, we find that contamination of the $\kcmb$ maps by tSZ signal necessitates removing a large fraction of the measurements. The specific choices of angular scale cuts are listed in Appendix \ref{sec:scalecuts}. While the bias coming from the tSZ effect can be reduced using data from multiple frequencies, the noise levels of the 95 and 220 GHz SPT-SZ data are such that this approach would not improve the overall signal-to-noise in the correlation functions..

We emphasize that the scale cuts imposed on $\galk$ and $\sheartk$ are not strictly necessary for the analysis of these two correlation functions.  Rather, these cuts were motivated by a desire to ensure that cosmological constraints from the analysis of the full \5x2pt data vector was not impacted by unmodeled effects.  The choice of angular scales to remove to eliminate biases in cosmological constraints is not unique: one can include more scales from a particular correlation function if scales are removed from another.  In this analysis, we have chosen to ensure consistency with the choices of the \3x2pt analysis of \citetalias{DESy1:2017} and the scale cut choices made therein.  Consequently, there is less tolerance for possible biases in the $\galk$ and $\sheartk$ correlation functions (see discussion in \citetalias{5x2methods}). This choice is reasonable if one views the primary purpose of this analysis as a consistency test of the \3x2pt results presented in \citetalias{DESy1:2017}. Furthermore, as shown in \citetalias{5x2methods} (Fig.~7), even without angular scale cuts, the total signal-to-noise of $\sheartk$ and $\galk$ is less than that of the \3x2pt analysis. The scale cut choices made here are therefore well motivated and do not result in a dramatic change to the \5x2pt cosmological constraints.

\subsection{Parameter constraints}
\label{sec:param_constraints}

We assume a Gaussian likelihood for modeling the observed data vectors. Given a data vector $\vec{d}$ representing some combination of two-point function measurements, and given the set of model parameters, $\vec{p}$, the data log-likelihood is:
\begin{equation}
\ln \mathcal{L}(\vec{d}|\vec{m}(\vec{p}))= 
-\frac{1}{2} \left( \vec{d} - \vec{m}(\vec{p}) \right)^T \mathbf{C}^{-1} \left( \vec{d} - \vec{m}(\vec{p}) \right),
\end{equation}
where $\vec{m}(\vec{p})$ is the model for the data vector described in Sec.~\ref{sec:model} and $\mathbf{C}$ is the covariance matrix. 

For the \3x2pt subset of observables, we compute the covariance between probes using an analytical, halo-model based covariance as described \cite{Krause:2017}. We extend the covariance estimate to include cross-correlations with $\kappa_{\rm CMB}$ as described in \citetalias{5x2methods}. 

While most of the contributions to the covariance matrix are calculated analytically, it was found in \citep{Troxel2018} that the geometry of the survey mask could impact the noise-noise term of the covariance significantly. For the DES \3x2pt block, this is corrected for by including the number of pairs in each angular bin when computing the correlation functions. For the two cross-correlation blocks that involve CMB lensing, the correction can not be applied trivially due to the scale-dependence in the noise spectrum. We therefore isolate this term in the analytically computed covariance matrix and replace it with a measurement using simulations. We generate 1000 Gaussian realizations\footnote{A large number of simulation realizations are needed to minimize the Anderson-Hartlap factor \citep{Hartlap2007}. } of CMB convergence noise, shape noise and random galaxy distributions with the same number density as data, and apply the mask that is used in the analysis.  We measure the $w^{\gammat\kcmb}(\theta)$ and $\galk$ from these for each realization, and compute the covariance using the ensemble, and add this covariance to the analytically computed component.

Given the data likelihood, the posterior on the model parameters, $P(\vec{p}|\vec{d})$, is calculated via
\begin{equation}
P(\vec{p}|\vec{d}) \propto \mathcal{L}(\vec{d} | \vec{p}) P_{\rm prior} (\vec{p}),
\end{equation}
where $P_{\rm prior}(\vec{p})$ is the prior on the model parameters. We adopt the same choice of parameter priors as in \citetalias{DESy1:2017} and use the \texttt{multinest} \citep{multinest} algorithm to sample the posterior distribution of the high-dimensional parameter space and to compute evidence integrals (see Sec.~\ref{sec:consistency_metrics}). The parameters explored in the analysis, as well as the priors used for each parameter are listed in Table~\ref{tab:params}. Since this work is primarily focused on examining consistency between \3x2pt and \5x2pt, we restrict our analysis to flat $\Lambda$CDM +$\nu$ cosmological models and will leave extensions to other models for future work.

\renewcommand{\arraystretch}{1.3}
\begin{table}
\caption{Parameters of the baseline model: fiducial values, flat priors (min, max), and Gaussian priors ($\mu$, $\sigma$).  Definitions of the parameters can be found in the text.  The cosmological model considered is spatially flat $\Lambda$CDM$+\nu$, so the curvature density parameter and equation of state of dark energy are fixed to $\Omega_{\mathrm K} = 0$ and $w=-1$, respectively.}
\begin{center}
\begin{tabular*}{0.33\textwidth}{@{\extracolsep{\fill}}| c c |}
\hline
\hline
Parameter &  Prior \\  
\hline 
\multicolumn{2}{|c|}{\textbf{Cosmology}} \\
$\om$  &  flat (0.1, 0.9)  \\ 
$A_\mathrm{s}/10^{-9}$ &  flat ($0.5$,$5.0$)  \\ 
$\ns$ & flat (0.87, 1.07)  \\
$\w$ &   fixed   \\
$\omb$  &  flat (0.03, 0.07)  \\
$h_0$   &  flat (0.55, 0.91)   \\
$\Omega_\nu h^2$ & flat$(5\times 10^{-4}, 10^{-2})$ \\
$\Omega_\mathrm{K}$ & 0 \\
$\tau$  & 0.08 \\
\hline
\multicolumn{2}{|c|}{\textbf{Galaxy bias}} \\
$b_\mr{g}^{i}$   & flat (0.8, 3.0) \\
\hline
\multicolumn{2}{|c|}{\textbf{Lens photo-z bias}} \\
$\Delta_{z,{\rm g}}^1 $  & Gauss (0.0, 0.007) \\
$\Delta_{z,{\rm g}}^2 $  & Gauss (0.0, 0.007) \\
$\Delta_{z,{\rm g}}^3 $  & Gauss (0.0, 0.006) \\
$\Delta_{z,{\rm g}}^4 $ & Gauss (0.0, 0.01) \\
$\Delta_{z,{\rm g}}^5 $  & Gauss (0.0, 0.01) \\
\hline
\multicolumn{2}{|c|}{\textbf{Source photo-z bias}} \\
$\Delta_{z,{\rm s}}^1 $  & Gauss (-0.001,0.016) \\
$\Delta_{z,{\rm s}}^2 $  & Gauss (-0.019,0.013) \\
$\Delta_{z,{\rm s}}^3 $  & Gauss (0.009, 0.011) \\
$\Delta_{z,{\rm s}}^4 $  & Gauss (-0.018, 0.022) \\
\hline
\multicolumn{2}{|c|}{\textbf{Shear Calibration bias }} \\
$m^i $ & Gauss (0.012, 0.023)\\
\hline
\multicolumn{2}{|c|}{\textbf{Intrinsic Alignments}} \\
$A_{\mathrm{IA,0}} $  & flat (-5.0, 5.0)\\
$\eta_{\mathrm{IA}} $  & flat (-5.0, 5.0)\\
$z_0$ & 0.62\\
\hline
\end{tabular*}
\end{center}
\label{tab:params}
\end{table}
\renewcommand{\arraystretch}{1.0}

\section{Data and Measurements}
\label{sec:data}

\subsection{DES-Y1 data}
The Dark Energy Survey (DES; \cite{DES2005}) is an optical imaging survey that covers 5000 deg$^{2}$ with 5 filter bands ($grizY$). The data is taken using the Dark Energy Camera \citep[DECam;][]{Flaugher2015} at the 4$\rm m$ Blanco telescope at the Cerro Tololo Inter-American Observatory. The first year data from DES was taken during the period Aug 2013 to Dec 2014, and covers roughly 1500 deg$^{2}$ to a median $10\sigma$ depth of $i\sim 22.9$. Approximately 1300 deg$^{2}$ of the Y1 data overlaps with the footprint of the SPT-SZ survey, and is the basis of the DES data in this analysis. An overview of available DES-Y1 data products can be found in \cite{Drlica-Wagner17}, while specific samples extracted for cosmological analyses are described in the individual two-point measurement papers \citep{ElvinPoole2017,Prat2017,Troxel2017}. The same galaxy and shape catalogs used in those papers are used for measuring cross-correlations with CMB lensing here.

\subsubsection{Tracer galaxies}
We use $\redmagic$-selected galaxies for the measurement of galaxy over-density, $\delta_{g}$. $\redmagic$\ is a sample of Luminous Red Galaxies (LRGs) generated using an algorithm that selects galaxies with reliable photometric redshifts \citep{Rozo2016redmagic}, and has photo-$z$ uncertainty at the level of $\sigma_{z}=0.017(1+z)$ \citep{ElvinPoole2017}. The redshift distributions were validated in \cite{Cawthon2017} by cross-correlating with spectroscopic samples.

The $\redmagic$\ samples are constructed so as to be volume-limited. Three catalogs with different luminosity cuts $L_{\rm min}$ (which results in different co-moving number densities) are used in this work: $L_{\rm min}=0.5L^{*}$ was used in the three lower redshift bins, whereas $L_{\rm min} = L^{*}$ and $L_{\rm min} = 1.5L^{*}$ were used in the two higher redshift bins. The five redshift bins are defined by $0.15<z<0.3$, $0.3<z<0.45$, $0.45<z<0.6$, $0.6<z<0.75$, and $0.75<z<0.9$. Ref.~\cite{ElvinPoole2017} subjected the $\redmagic$\ catalog to tests for systematic contamination from e.g. depth and seeing variation across the survey region. After weights are applied to the $\redmagic$\ galaxies to account for correlations between galaxy density and observational systematics (see \cite{ElvinPoole2017} for details), no evidence for significant residual contamination of the correlation function measurements is found across the range of angular scales considered. In addition to the tests on the galaxy catalogs that have gone into the \3x2pt analysis, we have performed in \citetalias{NKpaper} additional systematics tests specific to the cross-correlations with CMB lensing.

\subsubsection{Source galaxies}
We use the \textsc{MetaCalibration} \citep{Huff17,sheldon17} shear catalog for the background source galaxy shapes from which we extract $\gamma$.  \textsc{MetaCalibration} uses the data itself to calibrate shear estimates by artificially shearing the galaxy images and re-measuring the shear to determine the response of the shear estimator.  As in \citetalias{DESy1:2017}, the shear catalogs were divided into 4 tomographic bins: $0.2<z<0.43$, $0.43<z<0.63$, $0.63<z<0.9$, $0.9<z<1.3$, where $z$ is the mean of the redshift PDF for each galaxy as estimated from a modified version of the Bayesian Photometric Redshifts (BPZ) algorithm \citep{Benitez2000,Hoyle2017}. Descriptions of the shear catalog and the associated photo-$z$ catalog can be found in \cite{Zuntz17ca} and \cite{Hoyle2017}, respectively. In addition to the tests on the shear catalogs that were part of the \3x2pt analysis, we have performed additional systematics tests specific to the two-point correlation function $\sheartk$ in \citetalias{GKpaper}. 

\subsection{SPT-SZ and \Planck data}
We use the CMB lensing convergence map presented in \citetalias{Omori:2017} in this analysis. The \citetalias{Omori:2017} map is produced from an inverse-variance weighted linear combination of the SPT-SZ survey 150 GHz map and the \Planck 143 GHz map. Prior to combining these two maps, galaxy clusters detected with a signal-to-noise ratio ${\rm S}/{\rm N} \geq 6$ in \cite{Bleem2015} are masked with an aperture of radius $r=5'$. Point sources detected above 50 mJy (500 mJy) are masked with an aperture of radius $r=6'\ (r=9')$, while sources in the flux density range $6.4<F_{150}<50$~mJy are inpainted using the Gaussian constrained inpainting method. As shown in \citetalias{5x2methods}, our masking choices do not result in significant bias to the measured correlations with $\kappa_{\rm CMB}$.

The lensing map is reconstructed from the combined temperature map using the quadratic estimator of \cite{Okamoto2003}. The map is filtered to avoid noise and systematic biases from astrophysical foregrounds as described in \citetalias{Omori:2017}. This process generates a filtered lensing potential map $\bar{\phi}$, which is then converted to convergence via:
\begin{equation}
\hat{\kappa}_{\rm CMB}=\frac{1}{2\mathcal{R}^{\phi\phi}}\ell(\ell+1)(\bar\phi-\bar\phi^{\rm MF}),
\end{equation}
where $\mathcal{R}^{\phi\phi}$ is the response function, which is a multiplicative factor that renormalizes the filtered amplitude, and $\bar\phi^{\rm MF}$ is the mean-field bias, which we calculate by taking the average of simulated $\bar \phi$ maps. The response function is obtained by taking the ratio between the cross-correlation of input true $\phi$ and output reconstructed $\phi$ maps and the auto-correlation of input $\phi$. We remove modes with $\ell<30$ and $\ell>3000$ in the resulting convergence map. A Gaussian beam of $5.4'$ is then applied to the map to taper off the noise spectrum at high $\ell$.  We model the impact of this filtering using the $B(\ell)$ factor described in Sec.~\ref{sec:basic_model}.

When computing the correlation functions, we additionally apply a mask of radius $5'$ to clusters detected at signal-to-noise ${\rm S/N}>5$ in the SPT CMB maps \citep{Bleem2015} as well as DES $\redmapper$ clusters with richness $\lambda>80$ to further suppress contamination due to the tSZ effect.  In principle, such masking could bias the inferred correlation signals since clusters are associated with regions of large $\kappa_{\rm CMB}$; however, in \citetalias{5x2methods} we have quantified this effect and found it to be negligible.

\subsection{Measurements}
\label{sec:measurements}

The methodology for measuring $\galshear$, $\galgal$, $\shearshear$, $\sheartk$, and $\galk$ were presented in \cite{Prat2017},  \cite{ElvinPoole2017}, \cite{Troxel2017}, \citetalias{GKpaper}, and \citetalias{NKpaper}, respectively. All position-space correlation function measurements are carried out using the fast tree-code \textsc{TreeCorr}\footnote{\url{https://github.com/rmjarvis/TreeCorr}}.

\section{Consistency Metrics}  
\label{sec:consistency_metrics}
  
\citetalias{DESy1:2017} found cosmological constraints that were consistent with the $\Lambda$CDM cosmological model. One of the primary purposes of the present work is to perform a consistency test between the \3x2pt measurements and $\galk$+$\sheartk$ in the context of $\Lambda$CDM. Any evidence for inconsistency could indicate the presence of unknown systematics, or a breakdown in the $\Lambda$CDM model. For the purposes of assessing consistency, we use two different statistical metrics: the evidence ratio and the posterior predictive distribution. We describe these two approaches below. 
  
\subsection{Evidence ratio}

Several recent analyses, including \citetalias{DESy1:2017}, have used evidence ratios for quantifying consistency between different cosmological measurements. In this approach, consistency between different datasets is posed as a model selection problem. One effectively answers the question: "are the observations from two experiments more likely to be explained by a single (consistent) set of parameters, or by two different sets of parameters?"  If the datasets are more likely to be explained by a single set of model parameters, that can be interpreted as evidence for consistency between the measurements; alternatively, if the data are better explained by two different sets of parameters, that can be interpreted as evidence for inconsistency. 

To answer the question posed above, we use a ratio of evidences between two models, as motivated by \cite{Marshall:2006}. The Bayesian evidence (or marginal likelihood) for data, $D$, given a model $M$ and prior information $I$ is 
\begin{equation}
P(D | M, I) = \int \, P(D| \vartheta, M, I) P(\vartheta | M, I) d\vartheta,
\end{equation}
where $\vartheta$ represents the parameters of $M$. The quantity $P(D| \vartheta, M, I)$ represents the data likelihood, while $P(\vartheta | M, I)$ represents our prior knowledge of the parameters $\vartheta$.  

We would like to evaluate the consistency of two datasets, $D_1$ and $D_2$, under a cosmological model such as $\Lambda$CDM. Following \cite{Marshall:2006}, we introduce two models: $M_{\rm A}$  (which we will call the 
{\it consistency} model) and $M_{\rm B}$ (which we call the {\it inconsistency} model). In $M_{\rm A}$, the two datasets are described by a single set of model parameters.  In $M_{\rm B}$, on the other hand, there are two sets of model parameters, one describing $D_1$ and one describing $D_2$. The evidence ratio 
\begin{equation}
R \equiv \frac{P(D_1, D_2 | M_{\rm A}, I)}{P(D_1, D_2 | M_{\rm B}, I)},
\end{equation}
then provides a measure of the consistency between the two datasets. If $D_1$ and $D_2$ are independent, then the denominator can be written as a product of two evidences:
\begin{eqnarray}
\label{eq:evidence_ratio}
R &=& \frac{P(D_1, D_2 | M_{\rm A}, I)}{P(D_1, | M_{\rm B}, I)P(D_2, | M_{\rm B}, I)}  \\
&=& \frac{\int P(D_1, D_2| \vartheta, M, I) P(\vartheta | M, I) d \vartheta}{\int  P(D_1| \vartheta, M, I) P(\vartheta | M, I)d \vartheta \,\int  P(D_2| \vartheta, M, I) P(\vartheta | M, I)d \vartheta }, \nonumber \\
\end{eqnarray}
where in the second line, we have used $M$ to represent the cosmological model under which the consistency test is being performed, i.e. flat $\Lambda$CDM. In the \citetalias{DESy1:2017} approach, $M_{\rm B}$ assumes that $D_1$ and $D_2$ are independent, allowing us to simplify the evidence ratio as in Eq.~\ref{eq:evidence_ratio}. Additionally, the \citetalias{DESy1:2017} approach is to duplicate {\it all} of the model parameters when creating model $M_{\rm B}$, not just the cosmological parameters. We compute the multidimensional integrals in Eq.~\ref{eq:evidence_ratio} using \texttt{multinest}.   

In the evidence ratio approach, we interpret large $R$ as evidence for consistency between the datasets $D_1$ and $D_2$. It is common to use the Jeffreys scale \citep{Jeffreys1961} to assess the value of $R$.  An evidence ratio $\log_{10} R > 1$ would represent strong support for the consistency model; $\log_{10} R < 0$ would indicate preference for the inconsistency model. In the analysis of \citetalias{DESy1:2017}, a criterion of $\log_{10} R > -1$ was used as a threshold for combining datasets; an evidence ratio lower than this would indicate that the inconsistency model was preferred strongly enough that generating combined constraints from the two experiments would not be defensible.

The main advantage of the evidence ratio approach is that it is fully Bayesian, taking into account the full posterior on the model parameters, not just the maximum likelihood point (as is the case for e.g. a $\chi^2$ comparison).   However, the evidence ratio approach also has some significant drawbacks. For one, the alternative model considered in the evidence ratio (i.e. what we call $M_{\rm B}$ above) is not very well motivated. We have no a priori reason to think that doubling all of the parameters of $\Lambda$CDM and the systematics parameters provides a reasonable alternative model.  Furthermore, because the alternative model has so many parameters, it suffers a large Occam's razor penalty (see e.g. \cite{Raveri:2018}).  Finally, the assumption of independence between $D_1$ and $D_2$ built into model $M_{\rm B}$ is questionable when applied to the joint two-point function analyses, since we know that the measurements are indeed correlated from e.g. jackknife tests on the data. These concerns motivate us to explore alternative consistency metrics.

\subsection{Posterior predictive distribution}
\label{sec:ppd}

The posterior predictive distribution (PPD) provides an alternative (and still fully Bayesian) approach for evaluating consistency between two datasets, $D_1$ and $D_2$. Briefly, one generates plausible simulated realizations of $D_2$ given a posterior on the model parameters from $D_1$; we then ask whether these realizations look like the actual observed data, $D_2$.  The PPD approach has a long history in Bayesian analysis (for an overview of the method, see \cite{Gelman:2013}); it has recently been applied to cosmological analyses by e.g. \cite{Feeney:2018, Nicola:2018}.  

The PPD approach has a few advantages over the evidence ratio approach for assessing consistency between datasets. For one, the PPD addresses the question of consistency between datasets in the context of a {\it single} model.  In contrast, the evidence ratio approach poses the question of data consistency as a model comparison, and therefore requires proposal of an alternate model, which may not be well motivated.

Additionally, the PPD is computationally easy to implement since its main requirement is a set of parameter draws from a model posterior; this is easily generated with Markov chain Monte Carlo methods. On the other hand, the evidence ratio approach requires computing the Bayesian evidence; this can in principle be computed from parameter chains, but doing so is non-trivial and benefits from specialized algorithms like \texttt{multinest}. One potential drawback of the PPD approach is that it requires making a choice for how to compare the true data and the simulated realizations of the data.  Typically, to reduce the dimensionality of the problem, a test statistic is used for this comparison, such as the mean or $\chi^2$ (we will use $\chi^2$ below). However, a poor choice of test statistic will result in a less powerful test. We now summarize the computation of the PPD and its application to consistency tests of the measured two-point correlation functions.

We wish to determine whether the measurements of some new data vector, $D_2$, are reasonable given the posterior, $P(\vartheta | D_1, M, I)$, on model parameters, $\vartheta$, from the analysis of a different data vector, $D_1$, in the context of some model $M$ and given prior information $I$. For instance, below we will assess whether the observed $\galk$ and $\sheartk$ data vectors are reasonable given the \3x2pt constraints on flat $\Lambda$CDM.  To do this, we will generate simulated realizations of $D_2$, which we call $D_2^{\rm sim}$, conditioned on the posterior $P(\vartheta|D_1, M, I)$. The distribution of the simulated  realizations is
\begin{equation}
\label{eq:ppd_integral}
P(D^{\rm sim}_2 | D_1, M, I) = \int d\vartheta P(D^{\rm sim}_2 | \vartheta, D_1, M, I) P(\vartheta | D_{1},M,I).
\end{equation}
Note that we have allowed for the possibility that $D^{\rm sim}_{2}$ depends on  both $\vartheta$ and $D_{1}$.  This is important because when we use the PPD to determine consistency between $\sheartk$ and $\galk$ with \3x2pt, these data vectors are indeed correlated.  In practice, rather than computing the integral above, we generate many random realizations of $D_2^{\rm sim}$; this is the posterior predictive distribution.  

The distribution $P(D_2^{\rm sim} | \vartheta, D_1, M, I)$ can be computed given the likelihoods for $D_1$ and $D_2$. For the data vectors considered here, $(D_2, D_1)$ is distributed as a multivariate Gaussian:
\begin{equation}
(D_2,D_1) \sim \mathcal{N}\left( \left(\mu_2, \mu_1\right),\begin{bmatrix}
\Sigma_{22} & \Sigma_{21} \\
\Sigma_{12} & \Sigma_{11},
\end{bmatrix} \right), 
\end{equation}
where $\mu_{i}$ is the mean of $D_{i}$'s Gaussian distribution, and $\Sigma_{ij}$ represents the covariance between $D_{i}$ and $D_{j}$.  The distribution of $D_2$ conditioned on $D_1$ is then also a multivariate Gaussian:
\begin{equation}
D_2 | D_1 \sim \mathcal{N}\left( \mu_2 + \Sigma_{21}^{-1} \Sigma_{11}^{-1} \left(D_1 - \mu_1 \right), \Sigma_{22} - \Sigma_{21}\Sigma_{11}^{-1} \Sigma_{12} \right).
\end{equation}
Using these expressions, we can generate $D_2^{\rm sim}$ conditioned on $\vartheta$ and $D_1$, as in Eq.~\ref{eq:ppd_integral}.

To facilitate comparison between the true data, $D_2$, and the simulated draws, $D_2^{\rm sim}$, we define a test statistic, $T(D,\vartheta)$.  We will compute the distributions of both $T(D_2, \vartheta)$ and $T(D_2^{\rm sim}, \vartheta)$ (both of which depend on the posterior on $\vartheta$ from the analysis of $D_1$) and the comparison of these two distributions will be will used to assess consistency between $D_2$ and $D_1$.  Note that the test statistic can depend on $\vartheta$, but we are ultimately interested in the distribution of the test statistic marginalized over $\vartheta$, as in Eq.~\ref{eq:ppd_integral}.  Following \cite{Gelman:2013}, we choose $\chi^2$ as the test statistic, i.e. we set 
\begin{equation}
\label{eq:test_quantity}
T(D_2, \vartheta) = \left( D_2 - m(\vartheta) \right)^T \mathbf{C}^{-1} \left( D_2 - m(\vartheta) \right),
\end{equation}
where $m(\vartheta)$ is the model vector for $D_2$ generated using the parameters $\vartheta$, and $\mathbf{C}$ is the covariance of $D_2$. 

We compute the test statistics $T(D_2, \vartheta)$ and $T(D_2^{\rm sim}, \vartheta)$ for many $\vartheta$ drawn from the posterior $P(\vartheta | D_1, M, I)$.  At each $\vartheta$, $T(D_2^{\rm sim}, \vartheta)$  is computed by drawing a new $D_2^{\rm sim}$ from the distribution $P(D_2^{\rm sim} | \vartheta, D_1, M, I)$.  A $p$-value corresponding to the comparison between $T(D_2, \vartheta)$ and $T(D_2^{\rm sim}, \vartheta)$ is then computed as the fraction of the random draws for which $T(D_2^{\rm sim}, \vartheta) \geq T(D_2, \vartheta)$.  In other words, $p$ represents the probability that the simulated data has a higher test quantity than the observed real data.  A small $p$-value would indicate that the observed $D_2$ is unlikely given the posterior on model parameters from $D_1$, i.e. small $p$ would suggest inconsistency between $D_2$ and $D_1$.  A large $p$-value, on the other hand, could be an indication that that measurement uncertainty was overestimated.  Following standard practice, we will consider $p < 0.01$ or $p > 0.99$ to be cause for concern. 

\section{Blinding and Validation}
\label{sec:blinding}

To avoid possible confirmation bias during the analysis, the measurements and analyses of $\sheartk$ and $\galk$ were blinded until validation checks had passed. Based on projections from \citetalias{5x2methods}, we viewed the main purpose of the $\sheartk$ and $\galk$ measurements as a consistency check of the DES \3x2pt analysis. Consequently, we endeavored to blind ourselves to the consistency between the \3x2pt measurements with the additional $\sheartk$ and $\galk$.  

Blinding was implemented in several layers: first, the $\sheartk$ measurement was blinded by multiplying the shear values by some unknown amount, as described in \cite{Zuntz17ca}. Similarly, the $\galk$ data vector was multiplied by a random (unknown) number between 0.8 and 1.2. Second, two-point functions were never compared directly to theory predictions. Finally, cosmological contours were shifted to the origin or some other arbitrary point when plotting.

The following checks were required to pass before unblinding the measurements:
\begin{enumerate}
\item The shear and galaxy catalogs should pass all of the systematics tests described in \cite{Zuntz17ca} and \cite{ElvinPoole2017}.
\item The two-point function measurements should pass several systematic error tests: 
\begin{itemize}
\item Correlations between the cross-component of shear ($\gamma_{\times}$) and $\kcmb$ should be consistent with zero
\item The correlation of $\gamma$, $\delta_{\rm g}$ and $\kcmb$ with potential systematics maps should be consistent with zero or result in an acceptable small bias. More details regarding these tests are described in \citetalias{NKpaper} and \citetalias{GKpaper}.
\item Including prescriptions for effects not included in the baseline model (such as baryonic effects and nonlinear galaxy bias) should lead to acceptably small bias to cosmological constraints in simulated analyses (see \citetalias{5x2methods} for more details).
\end{itemize}
\item The covariance matrix estimate should be compared to data via jackknife estimates of the covariance matrix. See \citetalias{NKpaper} and \citetalias{GKpaper} for more details.
\end{enumerate}

Once the measurements were unblinded, we generated posterior samples from the joint analysis of $\galk$ and $\sheartk$. The measured data vectors were frozen at this point.

We required one final test before unblinding the cosmological constraints from the joint analysis of $\galk$ and $\sheartk$: the minimum $\chi^2$ from a flat $\Lambda$CDM fit to $\galk$ and $\sheartk$ should be less than some threshold value, $\chi^{2}_{\rm th}$. If the $\chi^2$ did not meet this threshold, it would indicate either a significant failure of flat $\Lambda$CDM model or an unidentified systematic; in either case, presenting constraints on $\Lambda$CDM would then be unjustified. The value of $\chi^{2}_{\rm th}$ was chosen such that the probability of getting a $\chi^2$ value so high by chance for a $\Lambda$CDM model, $p_{\rm th} \equiv P(\chi^2 > \chi^{2}_{\rm th}; \nu)$, was less than 0.01 for $\nu$ degrees of freedom. We determine in Appendix~\ref{sec:dof} that the effective number of degrees of freedom in this analysis is roughly 37.5. For $\nu = 37.5$, our choice of $p_{\rm th}$ corresponds to $\chi^2_{\rm th} = 60.5$. When fitting the model to the unblinded data, we found $\chi^2 = 32.2$. Since this was well below the threshold for unblinding, we proceeded to examining the $\Lambda$CDM posteriors and evaluating consistency with \3x2pt.  
  
\section{Results}
\label{sec:results}

\begin{figure}
\begin{center}
\includegraphics[height=0.45\textwidth]
{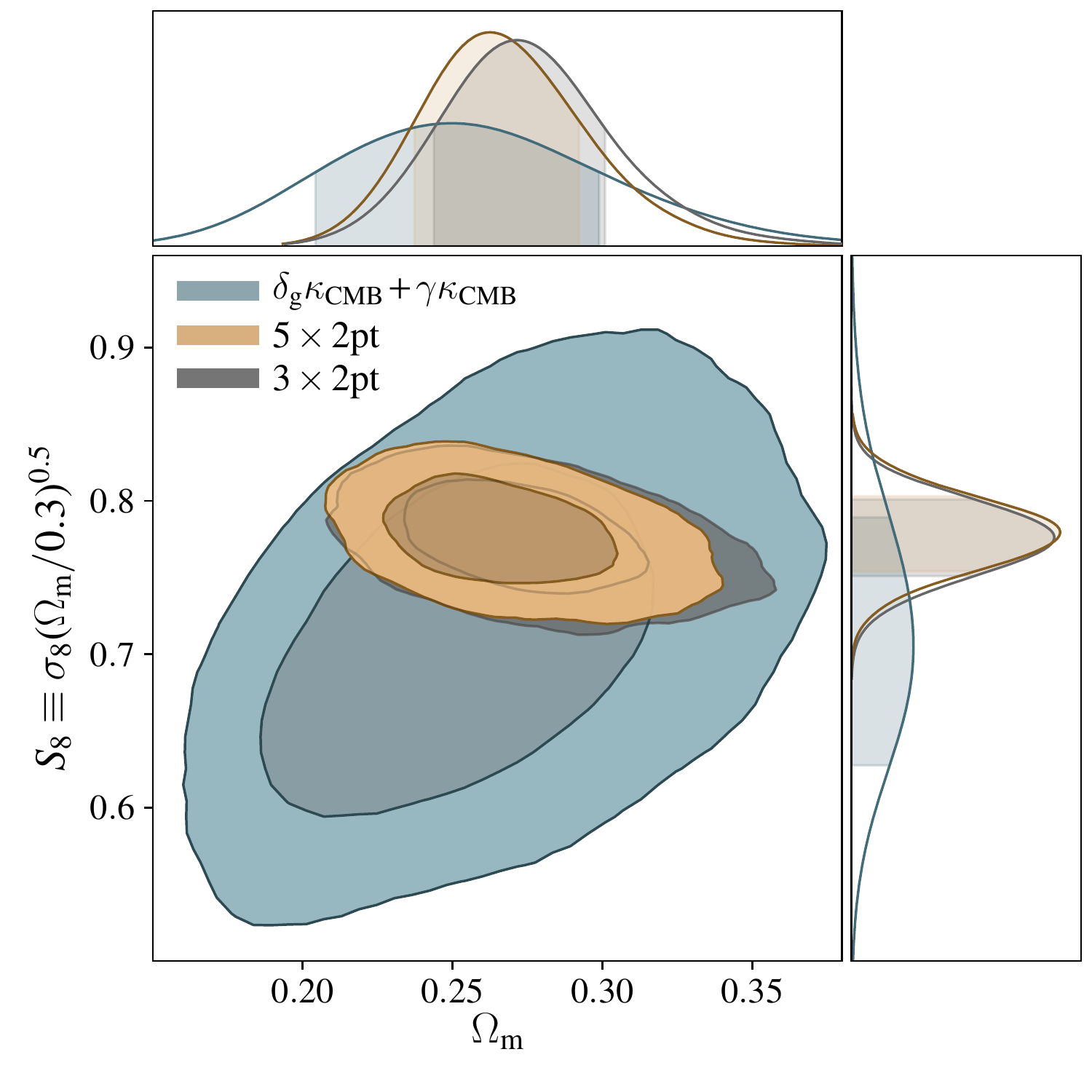}
\caption{Marginalized constraints on $\Omega_{\rm m}$ and $S_{8}\equiv \sigma_{8}(\Omega_{\rm m}/0.3)^{0.5}$ for different combinations of correlation functions in the context of $\Lambda$CDM+$\nu$ cosmology: \3x2pt (gray), $\sheartk + \galk$ (blue) and \5x2pt (gold). We note that the $\sheartk + \galk$ constraints have a different degeneracy direction compared to those of \3x2pt.}
\label{fig:fiducial_results}
\end{center}
\end{figure}

\subsection{Cosmological constraints from joint analysis of $\galk$ and $\sheartk$}

We first consider cosmological constraints from the joint analysis of $\galk$ and $\sheartk$.  The cosmological constraints obtained from $\sheartk$ alone were presented in \citetalias{GKpaper}. Similarly, cosmological constraints from the joint analysis of $\galk$ and $\wtheta$ were presented in \citetalias{NKpaper}.

Here, we focus on $\Omega_{\rm m}$ and $S_{8}$ as these parameters are tightly constrained in \3x2pt analysis, although constraints on several other parameters of interest are provided in Appendix \ref{sec:more_params}. From the joint analysis of $\galk$ and $\sheartk$, we obtain the constraints: 
\begin{eqnarray}
\Omega_{\rm m}&=& 0.250^{+0.040}_{-0.043} \nonumber\\
\sigma_{8}&=& 0.792^{+0.051}_{-0.096} \nonumber\\
S_{8} &\equiv& \sigma_{8}(\Omega_{\rm m}/0.3)^{0.5}=0.694^{+0.080}_{-0.059}.\nonumber
\end{eqnarray}
These constraints are shown as the blue contours in Fig.~\ref{fig:fiducial_results}.   Also overlaid in grey are the constraints from the \3x2pt analysis. While the signal-to-noise of $\galk+\sheartk$ is lower than that of \3x2pt, we observe that the degeneracy direction is complementary.  This suggests that once the constraining power of $\galk+\sheartk$ becomes more competitive, combining $\galk+\sheartk$ with \3x2pt could shrink the contours more efficiently due to degeneracy breaking. 

\subsection{Consistency between $\galk+\sheartk$ and \3x2pt}

The contours corresponding to $\galk+\sheartk$ (blue) and \3x2pt (grey) shown in Fig.~\ref{fig:fiducial_results} appear to be in good agreement. However, since projections of the high dimensional posterior (in this case 26 dimensional) into two dimensions can potentially hide tensions between the two constraints, we numerically assess tension between the two constraints using the two approaches described in Section~\ref{sec:consistency_metrics}. 

When evaluating consistency between $\galk+\sheartk$ and \3x2pt using the evidence ratio defined in Eq.~\ref{eq:evidence_ratio}, we find $\log_{10} R=2.3$. On the Jeffreys scale, this indicates "decisive" preference for the consistency model. This preference can be interpreted as evidence for consistency between the two data sets in the context of $\Lambda$CDM.

To use the PPD to assess consistency, we set $D_2$ equal to the combination of $\galk$ and $\sheartk$, and set $D_1$ equal to the \3x2pt data vector. Using the methods described in Sec.~\ref{sec:ppd}, we calculate $p = 0.48$ for this test, indicating that distribution of the test statistic $T(D,\vartheta)$ inferred from the measurements of $\galk$ and $\sheartk$ is statistically likely given the posterior on model parameters from the analysis of the \3x2pt data vector.  In other words, there is no evidence for inconsistency between $\galk$+$\sheartk$ and the \3x2pt measurements. The distribution of $T(D_2,\vartheta)$ and $T(D_2^{\rm sim},\vartheta)$ are shown in Fig.~\ref{fig:ppd_5x2} in the Appendix.  Consequently, both the evidence ratio metric and PPD metric indicate that there is no evidence for inconsistency between $\galk$+$\sheartk$ and the \3x2pt measurements in the context of flat $\Lambda$CDM. 

\subsection{\5x2pt constraints}
\label{sec:5x2constraints}

Since we find that the \3x2pt and $\galk$+$\sheartk$ measurements are not in tension using both the evidence ratio and PPD approaches, we now perform a joint analysis of the \3x2pt, $\galk$, and $\sheartk$ data vectors, i.e. the \5x2pt combination. Note that this analysis includes covariance between \3x2pt and the $(\galk, \sheartk)$ observables.

The cosmological constraints from this joint analysis are shown as gold contours in Fig.~\ref{fig:fiducial_results} (constraints on more parameters can be found in Section~\ref{sec:more_params}).
The cosmological constraints resulting from the \5x2pt analysis are: 
\begin{eqnarray}
\Omega_{\rm m}&=& 0.260^{+0.029}_{-0.019} \nonumber\\
\sigma_{8}&=& 0.821^{+0.058}_{-0.045} \nonumber\\
S_{8} &=&0.782^{+0.019}_{-0.025}.\nonumber
\end{eqnarray} 
The improvement in constraints when moving from the \3x2pt to \5x2pt analysis is small when considering the marginalized constraints on parameters. In Fig.~\ref{fig:fiducial_results}, some tightening of the constraints can be seen at high $\Omega_{\rm m}$ and low $\sigma_{8}$.  Some additional improvements can be seen in Appendix \ref{sec:more_params}.  We note that the parameters $n_s$, $\Omega_{\rm b}$, $h_0$, $\Omega_{\nu} h^2$ are all prior dominated in both the \3x2pt and \5x2pt analyses. 

Next, we compare the constraints on the linear galaxy bias obtained from the analysis of \3x2pt and \5x2pt data vectors. These are summarized in Table~\ref{table:bestfitb}. Adding the cross-correlations with $\kappa_{\rm CMB}$ to the \3x2pt analysis can help to break degeneracies between galaxy bias and other parameters (see \cite{5x2methods} for an explicit example).  However, given the relatively low signal-to-noise of these cross-correlations, we find that improvements in the galaxy bias constraints are not significant.

To assess the improvement across the full cosmological parameter space, we compute the ratio of the square roots of the determinants of the parameter covariance matrices for the \3x2pt and \5x2pt analyses.  This quantity effectively provides an estimate of the parameter space volume allowed by the posterior.  When performing this test, we restrict our consideration to those parameters actually constrained in the analysis: $\Omega_{\rm m}$, $A_{\rm s}$, the galaxy bias parameters and $A_{\rm IA}$ (the amplitude in the intrinsic alignment model). In this parameter subspace, the square root of the determinant of the covariance matrix is reduced by 10\% when going from the \3x2pt to the \5x2pt analysis. 

\begin{figure}
\begin{center}
\includegraphics[width=0.45\textwidth]
{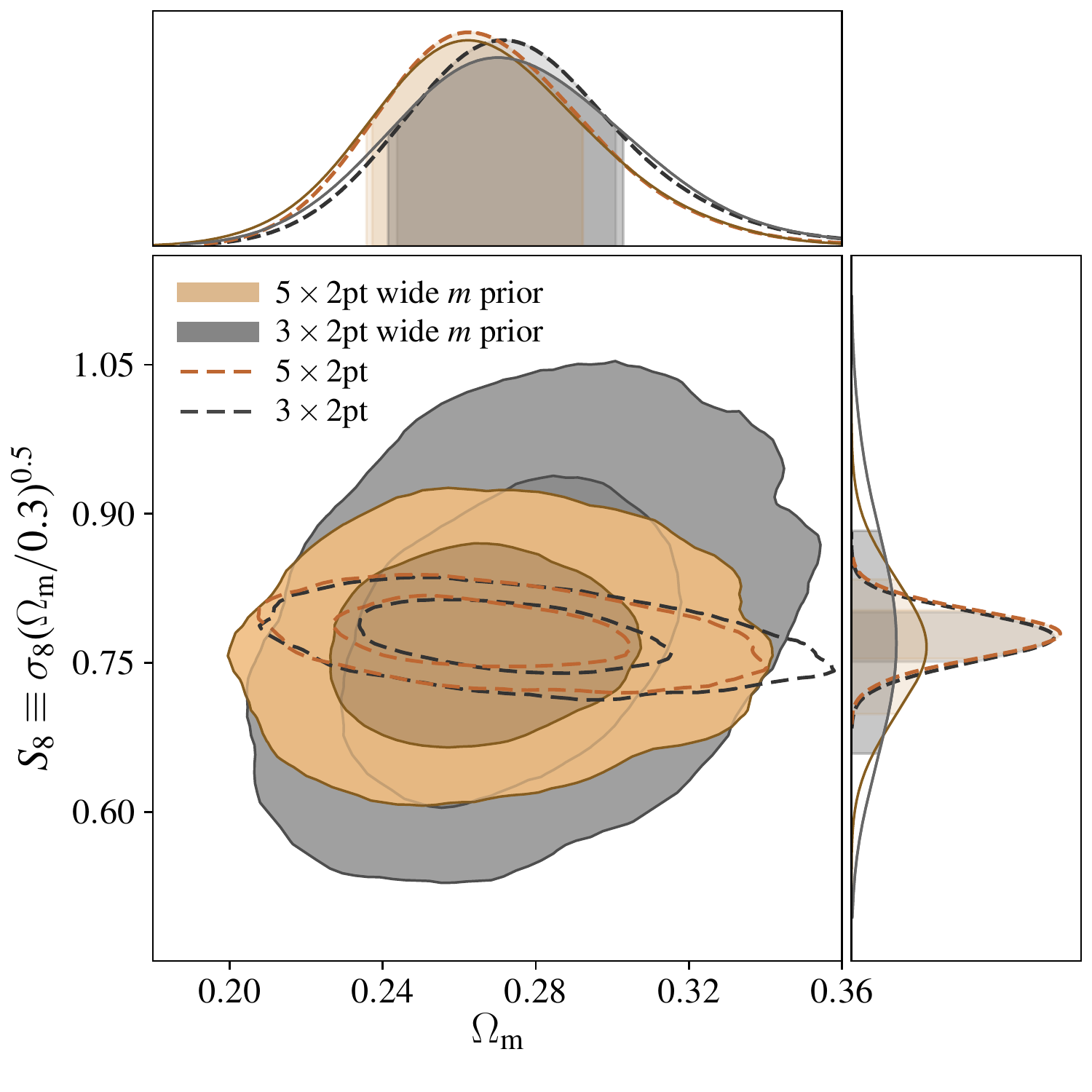}
\caption{Marginalized constraints on $\Omega_{\rm m}$ and $S_{8}\equiv \sigma_{8}(\Omega_{\rm m}/0.3)^{0.5}$ for the \3x2pt (gray) and \5x2pt (gold) combinations of correlation functions in the context of $\Lambda$CDM+$\nu$ cosmology when priors on multiplicative shear bias are relaxed (filled contours). In this case, the cosmological constraints obtained from the \5x2pt data vector are significantly tighter than those resulting from the \3x2pt data vector.  The dashed contours show the constraints when the fiducial priors on multiplicative shear bias (see Table~\ref{tab:params}) are applied. }
\label{fig:freem}
\end{center}
\end{figure}

\begin{table}
\renewcommand{\arraystretch}{1.5}
\centering
 \begin{tabular}{ccc}
\hline
\hline
\textsc{Sample} & \3x2pt $b_{i}$ & \5x2pt $b_{i}$\\
\hline
$0.15<z<0.30$  & $1.42^{+0.13}_{-0.08}$   & $1.41^{+0.11}_{-0.11}$\\
$0.30<z<0.45$  & $1.65^{+0.08}_{-0.12}$ & $1.60^{+0.11}_{-0.09}$\\
$0.45<z<0.60$  & $1.60^{+0.11}_{- 0.08}$  & $1.60^{+0.09}_{-0.10}$\\
$0.60<z<1.75$  & $1.93^{+0.14}_{- 0.10}$ & $1.91^{+0.11}_{-0.11}$\\
$0.75<z<1.90$  & $2.01^{+0.13}_{- 0.14}$ & $1.96^{+0.15}_{-0.11}$\\ \hline
\end{tabular}
\caption{Constraints on the linear galaxy bias parameters, $b_{i}$, from the \3x2pt and \5x2pt data vectors for the five redshift samples. }
\label{table:bestfitb}
\end{table}

\subsection{\5x2pt constraints with relaxed priors on multiplicative shear bias}

As shown in \citetalias{5x2methods}, one advantage of including cross-correlations with CMB lensing in a \5x2pt analysis is that these cross-correlations can help break degeneracies between the normalization of the matter power spectrum, galaxy bias, and multiplicative shear bias. For the fiducial DES-Y1 priors on multiplicative shear bias from \citetalias{DESy1:2017}, the degeneracy breaking is weak since multiplicative shear bias is already tightly constrained using data and simulation based methods, as described in  \cite{Zuntz17ca}. However, if these priors are relaxed, the \5x2pt analysis can obtain significantly tighter cosmological constraints than the \3x2pt analysis.  In essence, the cosmological constraints can be made more robust to the effects of multiplicative shear bias.

The \3x2pt and \5x2pt constraints on $\Omega_{\rm m}$ and $S_8$ when priors on multiplicative shear bias are relaxed to $m^i \in \left[-1,1\right]$ are shown in Fig.~\ref{fig:freem}.  In contrast to Fig.~\ref{fig:fiducial_results}, the \5x2pt constraints are significantly improved over \3x2pt when the multiplicative shear bias constraints are relaxed.  

For these relaxed priors, the data alone calibrate the multiplicative shear bias. The resultant constraints on the shear calibration parameters are shown in Table~\ref{table:bestfitm}. These constraints are consistent with the fiducial shear calibration priors shown in Table~\ref{tab:params}. In other words, we find no evidence for unaccounted systematics in DES measurements of galaxy shear. 

We have also performed similar tests for other nuisance parameters such as photo-$z$ bias and IA. However, the effect of self-calibration for these other parameters tends to be smaller than for shear calibration. As shown in \citetalias{5x2methods}, this is because shear calibration, galaxy bias, and $A_{\mathrm{s}}$ are part of a three-parameter degeneracy. Consequently, the \3x2pt data vector cannot tightly constrain these parameters without external priors on shear calibration. For the other systematics parameters, however, such strong degeneracies are not present, and significant self-calibration can occur.  Consequently, for these parameters, adding the additional correlations with $\kappa_{\rm CMB}$ does not add significant constraining power beyond that of the \3x2pt data vector.

\begin{table}
\renewcommand{\arraystretch}{1.5}
\centering
 \begin{tabular}{ccc}
\hline
\hline
\textsc{Sample} & \3x2pt $m_{i}$ & \5x2pt $m_{i}$\\
\hline
$0.20<z<0.43$  & $-0.03^{+0.34}_{-0.16}$   & $\hspace{0.2cm}0.03^{+0.25}_{-0.15}$\\
$0.43<z<0.63$  & ${-0.02}^{+0.27}_{-0.14}$ & $\hspace{0.2cm}0.07^{+0.19}_{-0.11}$\\
$0.63<z<0.90$  & ${-0.04}^{+0.20}_{- 0.15}$  & $-0.01^{+0.13}_{-0.09}$\\
$0.90<z<1.30$  & ${-0.02}^{+0.18}_{- 0.17}$ & $-0.08^{+0.14}_{-0.08}$\\
\hline
\end{tabular}
\caption{Constraints on the shear calibration parameters, $m_{i}$, from the \3x2pt and \5x2pt data vectors when priors on $m_{i}$ are relaxed.  In all cases, the posteriors obtained on the $m_i$ from the \5x2pt analysis are consistent with the priors adopted in the \3x2pt analysis of \cite{DESy1:2017}.}
\label{table:bestfitm}
\end{table}

\begin{figure}
\begin{center}
\includegraphics[height=0.45\textwidth]
{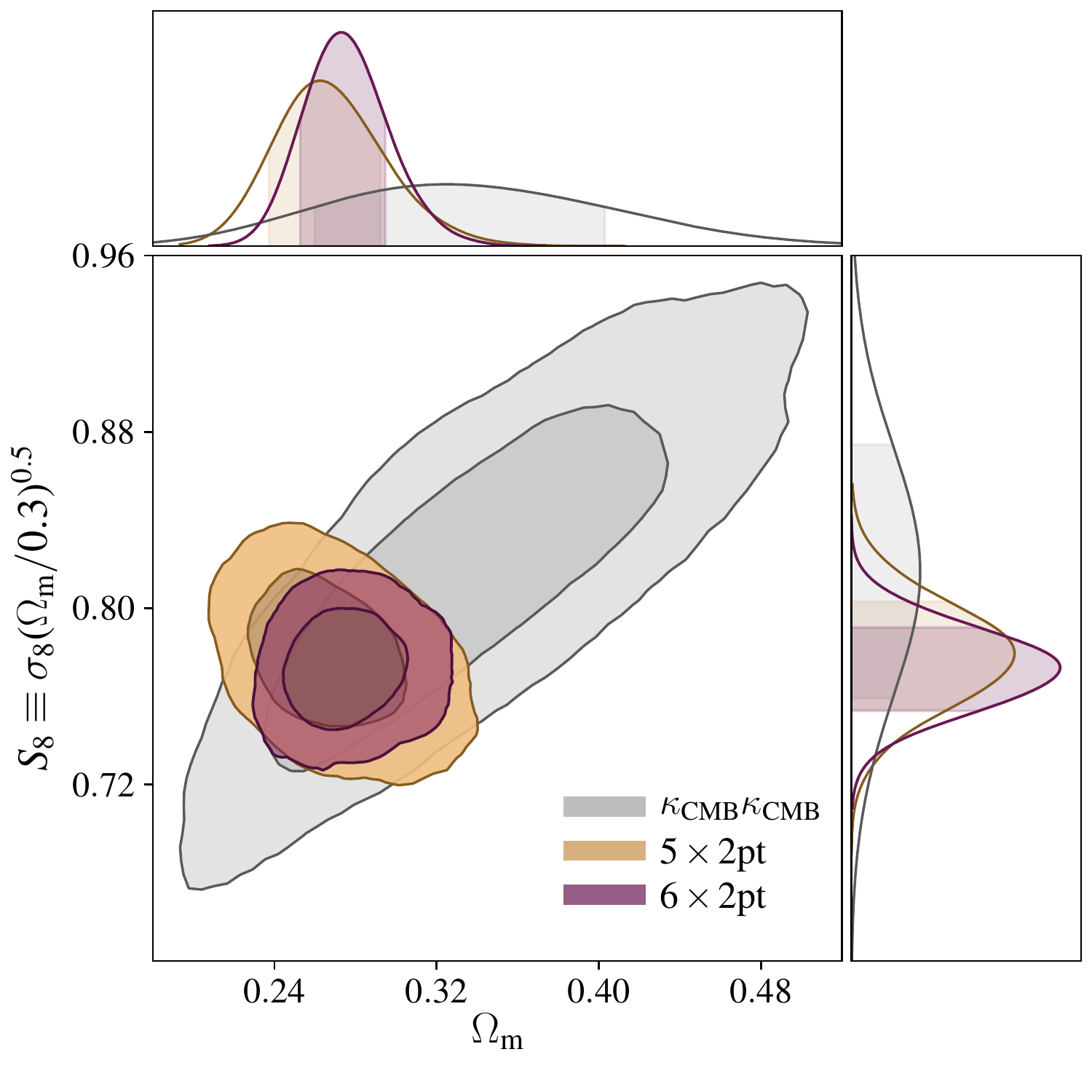}
\caption{Marginalized constraints on $\Omega_{\rm m}$ and $S_{8}\equiv \sigma_{8}(\Omega_{\rm m}/0.3)^{0.5}$ for different combinations of correlation functions in the context of $\Lambda$CDM+$\nu$ cosmology: \5x2pt (gold), $w^{\kappa_{\rm CMB} \kappa_{\rm CMB}}$ (gray) and \6x2pt (purple). The $w^{\kappa_{\rm CMB} \kappa_{\rm CMB}}$ contours are derived from the \Planck 2015 lensing data \citep{Planck:cmblensing}. The \5x2pt contours are identical to those in Fig.~\ref{fig:fiducial_results}. The $w^{\kappa_{\rm CMB} \kappa_{\rm CMB}}$ constraints are complementary to those of the \5x2pt analysis.}
\label{fig:6x2_results}
\end{center}
\end{figure}

\subsection{Consistency with \Planck measurements of the CMB lensing autospectrum}
\label{sec:kk_consistency}

While the \5x2pt data vector includes cross-correlations of galaxies and galaxy shears with CMB lensing, it does not include the CMB lensing auto-spectrum. Both the \5x2pt data vector and CMB lensing auto-spectrum are sensitive to the same physics, although the CMB lensing auto-spectrum is sensitive to higher redshifts as a result of the CMB lensing weight peaking at $z \sim 2$. Consistency between these two datasets is therefore a powerful test of the data and the assumptions of the cosmological model.  

Measurements of the CMB lensing autospectrum over the 2500 ${\rm deg}^{2}$ patch covered by the SPT-SZ survey have been obtained from a combination of SPT and \Planck data by \cite{Omori:2017}, and this power spectrum has been used to generate cosmological constraints by \cite{Simard2018}. Because of lower noise and higher resolution of the SPT maps relative to {\it Planck}, the cosmological constraints obtained in \cite{Simard2018} are comparable to those of the full sky measurements of the CMB lensing autospectrum presented in \cite{Planck:cmblensing}, despite the large difference in sky coverage.

In this analysis, we choose to test for consistency between the \5x2pt data vector and the {\it Planck}-only measurement of the CMB lensing autospectrum.  The primary motivation for this choice is that it significantly simplifies the analysis because it allows us to ignore covariance between the \5x2pt data vector and the CMB lensing autospectrum. This simplification comes at no reduction in cosmological constraining power. Furthermore, the SPT+{\it Planck} and {\it Planck}-only measurements of the CMB lensing autospectrum are consistent \citep{Simard2018}. 

Ignoring the covariance between the \5x2pt data vector and the {\it Planck} CMB lensing autospectrum measurements is justified for several reasons. First, the CMB lensing auto-spectrum is most sensitive to large scale structure at $z \sim 2$, at significantly higher redshifts than that probed by the \5x2pt data vector. Second, the instrumental noise in the SPT CMB temperature map is uncorrelated with noise in the \Planck CMB lensing maps. Finally, and most significantly, the measurements of the \5x2pt data vector presented here are derived from roughly 1300 ${\rm deg}^{2}$ of the sky, while the \Planck lensing autospectrum measurements are full-sky. Consequently, a large fraction of the signal and noise in the \Planck full-sky lensing measurements is uncorrelated with that of the \5x2pt data vector. We therefore treat the \Planck CMB lensing measurements as independent of the \5x2pt measurements in this analysis. 

The cosmological constraints from \Planck lensing autospectrum measurements alone are shown as the grey contours in Fig.~\ref{fig:6x2_results}. The constraints from the \5x2pt analysis and those of the \Planck lensing autospectrum overlap in this two dimensional projection of the multidimensional posteriors.  We find an evidence ratio of $\log_{10} R = 4.1$ when evaluating consistency between the \5x2pt data vector and the \Planck lensing autospectrum measurements, indicating ``decisive'' preference on the Jeffreys scale for the consistency model. 

When using the PPD to assess consistency, we set $D_2$ equal to $\kk$ and set $D_1$ equal to the \5x2pt data vector. The $p$-value computed from the PPD is determined to be $p=0.09$; there therefore no significant evidence for inconsistency between the \5x2pt and $w^{\kappa_{\rm CMB}\kappa_{\rm CMB}}$ measurements in the context of $\Lambda$CDM. The distributions of the test statistic for the data and realizations are shown in Fig.~\ref{fig:ppd_6x2} in the Appendix.

\subsection{Combined constraints from \5x2pt and the \Planck lensing autospectrum}
	
Having found that the cosmological constraints from the \5x2pt and \Planck lensing analyses are statistically consistent, we perform a joint analysis of both datasets, i.e. of the \6x2pt data vector. The constraints resulting from the analysis of this joint data vector are shown as the purple contours in Fig.~\ref{fig:6x2_results} (constraints on more parameters can be found in Section~\ref{sec:more_params}).

As seen in Fig.~\ref{fig:6x2_results}, the DES+SPT+\Planck \5x2pt analysis yields cosmological constraints that are complementary to the auto-spectrum of \Planck CMB lensing, as evidenced by the nearly orthogonal degeneracy directions of the two contours in $\Omega_{\rm m}$ and $S_8$. When combining the constraints, we obtain for the \6x2pt analysis:
\begin{eqnarray}
\Omega_{\rm m}&=& 0.271^{+0.022}_{-0.016}\nonumber\\
\sigma_{8}&=& 0.800^{+0.040}_{-0.025} \nonumber\\
S_{8}&=& 0.776^{+0.014}_{-0.021}.\nonumber
\end{eqnarray}
The constraints on $\Omega_{\rm m}$ and $S_8$ are 25\% and 24\% tighter, respectively, than those obtained from the \3x2pt analysis of \citetalias{DESy1:2017}. The addition of $\planck$ lensing provides additional constraining power coming from structure at higher redshifts than is probed by DES. 

\section{Discussion}
\label{sec:discussion}

We have presented a joint cosmological analysis of two-point correlation functions between galaxy density, galaxy shear and CMB lensing using data from DES, the SPT-SZ survey and \emph{Planck}. The \5x2pt observables --- $\wtheta$, $\xi_{\pm}(\theta)$, $\gglensing$, $\galk$, and $\sheartk$ --- are sensitive to both the geometry of the Universe and to the growth of structure out to redshift $z \lesssim 1.3$.\footnote{The cross-correlations with $\kappa_{\rm CMB}$ depend on the distance to the last scattering surface at $z \sim 1100$ through the lensing weight of Eq.~\ref{eq:qkcmb}. This sensitivity is purely geometric, though, and does not reflect sensitivity to large scale structure at high redshifts.} The measurement process and analysis has been carried out using a rigorous blinding scheme, with cosmological constraints being unblinded only after nearly all analysis choices were finalized and systematics checks had passed.

We have used two approaches --- one based on an evidence ratio and one based on the posterior predictive distribution --- to evaluate the consistency between constraints from $\galk$ and $\sheartk$ and those obtained from the \3x2pt data vector explored in \citetalias{DESy1:2017}. We find no evidence for tension between these two datasets in the context of flat $\Lambda$CDM+$\nu$ cosmological models. This is a powerful consistency test of the \citetalias{DESy1:2017} results given that the CMB lensing measurements rely on completely different datasets from the DES observables, and are subject to very different sources of systematic error. Since we find these datasets to be statistically consistent, we perform a joint analysis of the \5x2pt data vector to obtain cosmological constraints, with the results shown in Fig.~\ref{fig:fiducial_results}.   The reduction parameter volume in going from the \3x2pt to \5x2pt data vector, as measured by the square root of the determinant of the parameter covariance matrix, is roughly 10\% over the subspace of parameters most constrained by these analyses ($\Omega_{\rm m}$, $A_\mathrm{s}$, the galaxy bias parameters, and $A_{\rm IA}$).

Notably, when priors on the multiplicative shear bias parameters are relaxed, we find that the \5x2pt data vector yields significantly tighter cosmological constraints than the \3x2pt data vector (Fig.~\ref{fig:freem}).  The inclusion of the CMB lensing cross-correlations in the analysis allows the data to self-calibrate the shear bias parameters (although not at the level of the DES priors).

The autocorrelation of CMB lensing convergence is sensitive to a wide range of redshifts, with the CMB lensing weight peaking at $z \sim 2$, and having significant support from higher redshifts. Again using both evidence ratio and posterior predictive distribution-based tests, we evaluate the consistency between the \Planck CMB lensing autospectrum measurements and the \5x2pt combination of observables. We find the two data sets to be consistent in the context of flat $\Lambda$CDM cosmological models, justifying a joint analysis.  The constraints using the full \6x2pt combination of correlation functions are shown in Fig.~\ref{fig:6x2_results}.

%We note that in our fiducial analysis, the gain in constraining power in going from the DES-only combination of two-point observables (i.e. \3x2pt) to the combination including cross-correlation with CMB lensing (i.e. \5x2pt) is fairly small, as seen in Fig.~\ref{fig:fiducial_results}.   

To some extent, the small improvement in cosmological constraints between the \3x2pt and \5x2pt analysis is a consequence of our conservative angular scale cuts. As noted in Sec.~\ref{sec:scale_cuts}, the choice of angular scale cuts adopted here favors the \3x2pt analysis over the cross-correlations with CMB lensing. This choice was motivated by the desire to perform a consistency test of the \3x2pt results, but a different choice could significantly impact the relative strengths of the \3x2pt and \5x2pt analyses. Furthermore, while the scale cuts for the cross-correlations with CMB lensing were informed by consideration of several unmodeled effects \citepalias{5x2methods}, they were driven primarily by issues of tSZ bias in the $\kappa_{\rm CMB}$ maps of \cite{Omori:2017}. Future measurement with DES and SPT, however, have the potential to significantly improve the constraining power of CMB lensing cross-correlations and to dramatically alleviate the problem of tSZ bias. We describe some of these expected improvements in more detail below.

Data quality and volume are expected to improve significantly in the near future for several reasons. First, the current data uses only first year DES observations. With full survey, DES will cover roughly 5000 sq. deg. (relative to the $\sim 1300$ sq. deg. considered here) and will reach at least a magnitude deeper. On the SPT side, significantly deeper observations of the CMB over a 500 ${\rm deg}^{2}$ patch have been made by SPTpol \citep{SPTpol}. There are also somewhat shallower observations over another 2500 ${\rm deg}^{2}$ field that partly overlaps with the DES footprint; this field was observed with SPTpol when the sun was too close to the 500 ${\rm deg}^{2}$ field. Additionally, several ongoing and upcoming CMB experiments have significant overlap with the DES survey region, and should enable significantly higher signal-to-noise measurements of $\kappa_{\rm CMB}$ over this footprint. These include SPT-3G \citep{Benson:2014}, Advanced ACTPol \citep{Henderson:2016}, the Simons Observatory \citep{SimonsObs2018} and CMB Stage-4 \citep{Abazajian:2016}.

Measurement algorithms are also expected to improve significantly in the near future. On the DES side, better data processing and shear measurement algorithms will likely enable the use of lower signal-to-noise source galaxies, and lead to tighter priors on the shear calibration bias. Photometric redshift determination is also expected to improve with future DES analyses. On the SPT side, contamination of the $\kappa_{\rm CMB}$ maps can be significantly reduced using multifrequency component separation methods to remove tSZ. In combination with data from {\it Planck}, this cleaning process can be accomplished with little reduction in signal-to-noise using the method outlined in \cite{Madhavacheril:2018}. 

Finally, several improvements are expected on the modeling side. As discussed in \citetalias{Krause:2017} and \citetalias{5x2methods}, effects such as the impact of baryons on the matter power spectrum and nonlinear galaxy bias were ignored in this analysis. This model simplicity necessitated restriction of the data to the regime where these approximations were valid, removing a significant fraction of the available signal-to-noise. With efforts to improve modeling underway, we can expect to exploit more of the available signal-to-noise of the two-point measurements using future data from DES and SPT.

This work represents the joint analysis of six two-point functions of large scale structure, measured using three different cosmological surveys, and spanning redshifts from $z = 0$ to $z \sim 1100$. Remarkably, although the three observables considered in this work --- $\delta_{\rm g}$, $\gamma$ and $\kappa_{\rm CMB}$ --- are measured in completely different ways, the two-point correlation measurements are all consistent under the flat $\Lambda$CDM cosmological model. The combined constraints from these measurements of correlation functions of the large scale structure are some of the tightest cosmological constraints to date, and are highly competitive with other cosmological probes. With significant improvements to data and methodology expected in the near future, two-point functions of large scale structure will continue to be a powerful tool for studying our Universe.

\section*{Acknowledgements}

Funding for the DES Projects has been provided by the U.S. Department of Energy, the U.S. National Science Foundation, the Ministry of Science and Education of Spain, 
the Science and Technology Facilities Council of the United Kingdom, the Higher Education Funding Council for England, the National Center for Supercomputing 
Applications at the University of Illinois at Urbana-Champaign, the Kavli Institute of Cosmological Physics at the University of Chicago, 
the Center for Cosmology and Astro-Particle Physics at the Ohio State University,
the Mitchell Institute for Fundamental Physics and Astronomy at Texas A\&M University, Financiadora de Estudos e Projetos, 
Funda{\c c}{\~a}o Carlos Chagas Filho de Amparo {\`a} Pesquisa do Estado do Rio de Janeiro, Conselho Nacional de Desenvolvimento Cient{\'i}fico e Tecnol{\'o}gico and 
the Minist{\'e}rio da Ci{\^e}ncia, Tecnologia e Inova{\c c}{\~a}o, the Deutsche Forschungsgemeinschaft and the Collaborating Institutions in the Dark Energy Survey. 

The Collaborating Institutions are Argonne National Laboratory, the University of California at Santa Cruz, the University of Cambridge, Centro de Investigaciones Energ{\'e}ticas, 
Medioambientales y Tecnol{\'o}gicas-Madrid, the University of Chicago, University College London, the DES-Brazil Consortium, the University of Edinburgh, 
the Eidgen{\"o}ssische Technische Hochschule (ETH) Z{\"u}rich, 
Fermi National Accelerator Laboratory, the University of Illinois at Urbana-Champaign, the Institut de Ci{\`e}ncies de l'Espai (IEEC/CSIC), 
the Institut de F{\'i}sica d'Altes Energies, Lawrence Berkeley National Laboratory, the Ludwig-Maximilians Universit{\"a}t M{\"u}nchen and the associated Excellence Cluster Universe, 
the University of Michigan, the National Optical Astronomy Observatory, the University of Nottingham, The Ohio State University, the University of Pennsylvania, the University of Portsmouth, 
SLAC National Accelerator Laboratory, Stanford University, the University of Sussex, Texas A\&M University, and the OzDES Membership Consortium.

Based in part on observations at Cerro Tololo Inter-American Observatory, National Optical Astronomy Observatory, which is operated by the Association of 
Universities for Research in Astronomy (AURA) under a cooperative agreement with the National Science Foundation.

The DES data management system is supported by the National Science Foundation under Grant Numbers AST-1138766 and AST-1536171.
The DES participants from Spanish institutions are partially supported by MINECO under grants AYA2015-71825, ESP2015-66861, FPA2015-68048, SEV-2016-0588, SEV-2016-0597, and MDM-2015-0509, 
some of which include ERDF funds from the European Union. IFAE is partially funded by the CERCA program of the Generalitat de Catalunya.
Research leading to these results has received funding from the European Research
Council under the European Union's Seventh Framework Program (FP7/2007-2013) including ERC grant agreements 240672, 291329, and 306478.
We  acknowledge support from the Australian Research Council Centre of Excellence for All-sky Astrophysics (CAASTRO), through project number CE110001020, and the Brazilian Instituto Nacional de Ci\^encia
e Tecnologia (INCT) e-Universe (CNPq grant 465376/2014-2).

This manuscript has been authored by Fermi Research Alliance, LLC under Contract No. DE-AC02-07CH11359 with the U.S. Department of Energy, Office of Science, Office of High Energy Physics. The United States Government retains and the publisher, by accepting the article for publication, acknowledges that the United States Government retains a non-exclusive, paid-up, irrevocable, world-wide license to publish or reproduce the published form of this manuscript, or allow others to do so, for United States Government purposes.

The South Pole Telescope program is supported by the National Science Foundation through grant PLR-1248097. Partial support is also provided by the NSF Physics Frontier Center grant PHY-0114422 to the Kavli Institute of Cosmological Physics at the University of Chicago, the Kavli Foundation, and the Gordon and Betty Moore Foundation through Grant GBMF\#947 to the University of Chicago. The McGill authors acknowledge funding from the Natural Sciences and Engineering Research Council of Canada, Canadian Institute for Advanced Research, and Canada Research Chairs program.

This research used resources of the National Energy Research Scientific Computing Center (NERSC), a DOE Office of Science User Facility supported by the Office of Science of the U.S. Department of Energy under Contract No. DE-AC$02$-$05$CH$11231$.

\bibliography{thebibliography.bib}

\newpage

\appendix

\section{Scale Cuts}
\label{sec:scalecuts}

The minimum angular scales for each of the five correlation functions are listed below:
\begin{align}
\theta_{\min}^{\delg\delg} =&[43',27',20',16',14'], \notag \\
\theta_{\min}^{\delg\gamma} =&[64',40',30',24',21'],  \notag \\
\xi_{+} =&[7.2',7.2',5.7',5.7',7.2',4.5',4.5',4.5',5.7',4.5',3.6',3.6'];  \notag \\
\xi_{-} =&[90.6',72.0',72.0',72.0',72.0',57.2',57.2',45.4', \notag \\
&72.0',57.2',45.4',45.4',72.0',45.4',45.4',36.1'] \notag \\
\theta_{\min}^{\delg\kcmb}=&[15',25',25',15',15'],  \notag \\
\theta_{\min}^{\gamma\kcmb}=&[40',40',60',60'].
\end{align}
The 5 (4) values correspond to the 5 (4) redshift bins for the $\delta_{g}$ ($\gamma$) fields in the cross-correlations $\delta_{g}\delta_{g}$, $\delta_{g}\gamma$, $\delg\kcmb$ and $\gamma\kcmb$. For $\xi_{\pm}$, the values correspond to all the auto and cross-correlations between the 4 source redshift bins (the numbers are ordered as bin1-bin1, bin1-bin2, ... bin2-bin1, bin2-bin2...). 

\begin{figure*}
\includegraphics[width=0.9\textwidth]
{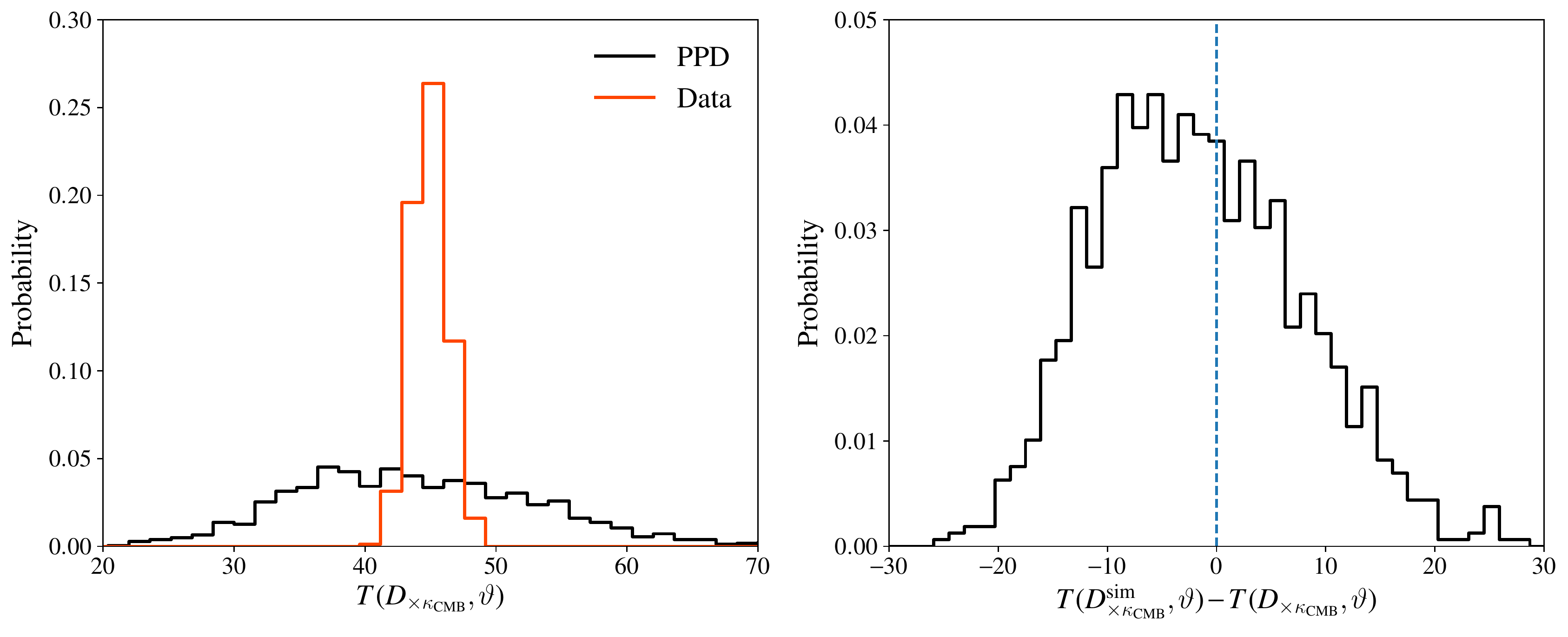}
\caption{Results of the PPD analysis to assess consistency between the \3x2pt and the $D_{\times \kappa_{\rm CMB}} \equiv \left( 
\sheartk, \galk \right)$ data vector in the context of flat $\Lambda$CDM$+\nu$.  Left panel shows the distributions of test quantities, $T(D,\vartheta)$, for the simulated $D_{\times \kappa_{\rm CMB}}$ data vector (black) and real data vector (orange), given the posterior on model parameters from the analysis of the \3x2pt data vector.  The test quantity used in this analysis is defined in Eq.~\ref{eq:test_quantity}.  Right panel shows the distribution of the difference between the test quantity evaluated on the PPD realizations and on the real data, $T(D_2^{\rm sim},\vartheta) - T(D_2,\vartheta)$.  Given the posterior on the model parameters from the \3x2pt data vector, approximately 44\% of the simulated realizations of $\sheartk$ and $\galk$ lead to a test quantity that is greater than that of the actual data.  This indicates that $D_{\times \kappa_{\rm CMB}}$ is statistically consistent with the \3x2pt data vector under the flat $\Lambda$CDM model.}
\label{fig:ppd_5x2}
\end{figure*}

\begin{figure*}
\includegraphics[width=0.9\textwidth]
{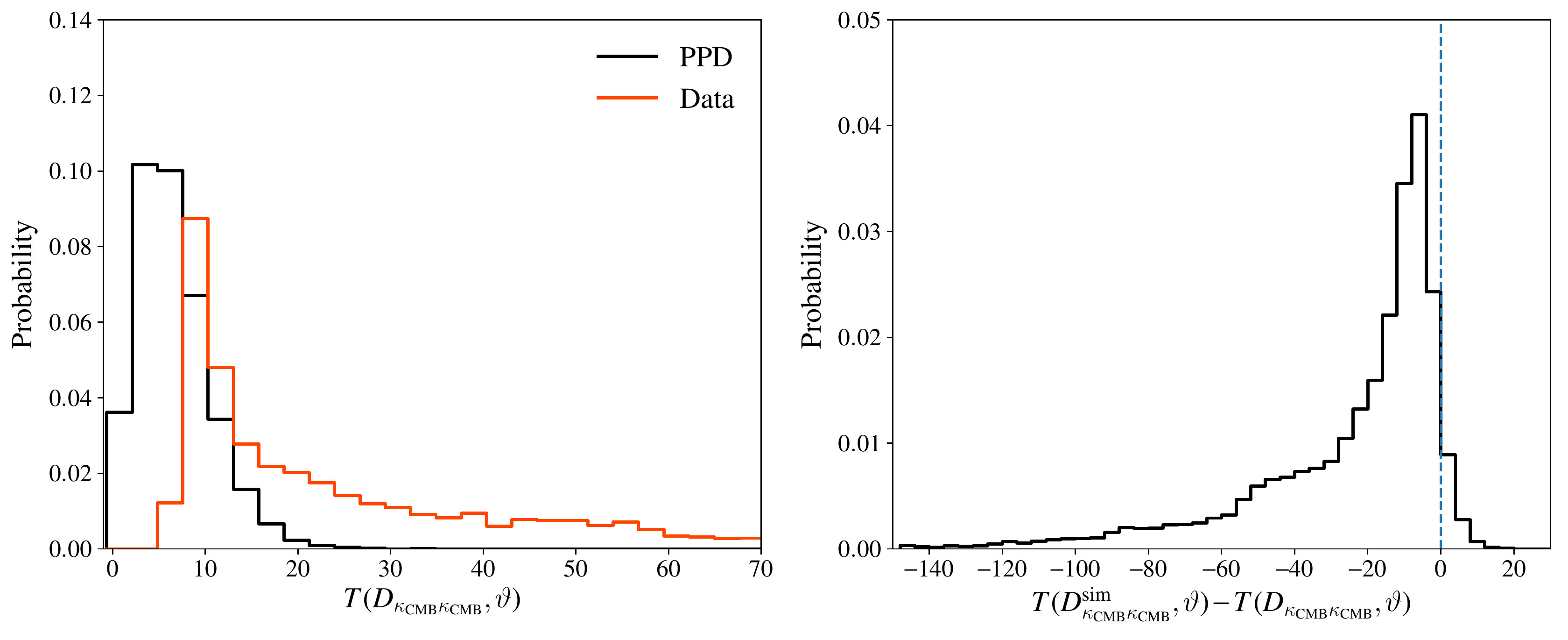}
\caption{Same as Fig.~\ref{fig:ppd_5x2}, except now showing results of PPD analysis to assess consistency between \5x2pt data vector and $\kk$ data vector.  In this case, approximately 9\% of the simulated realizations of $\kk$ lead to a test quantity that is greater than that of the actual data.  This indicates that $\kk$ is statistically consistent with the \5x2pt data vector under the flat $\Lambda$CDM model.}
\label{fig:ppd_6x2}
\end{figure*}

\section{Finding the DOF}
\label{sec:dof}
Determining the appropriate counting of degrees of freedom, $\nu$, to use when performing the $\chi^2$ test of the flat $\Lambda$CDM fit to the $\galk$ and $\sheartk$ data vector is complicated by the fact that the effects of some of the parameters in the model may be partially degenerate, and by the fact that we impose informative priors on some parameters.  We determined an effective $\nu$ by generating many simulated noisy data vectors from the theory model and covariance matrix, and fitting these to determine the minimum value of $\chi^2$.  We then fit these simulated data vectors to determine, $\chi^2_{\rm min,i}$, the minimum values of $\chi^2$ for the $i$th data vector.  Finally, we fit the distribution of $\chi^2_{\rm min,i}$ to a $\chi^2$ distribution to extract a constraint on $\nu$.  We find $\nu = 37.5 \pm 1.7$.  

\section{Posterior predictive distributions}

As discussed in the main text, we assess consistency between \3x2pt and $(\galk, \sheartk)$, and between \5x2pt and $w^{\kappa_{\rm CMB}\kappa_{\rm CMB}}$ using both evidence ratio and PPD-based approaches.  In this appendix, we show the distributions of the PPD test statistic.

Fig.~\ref{fig:ppd_5x2} and Fig.~\ref{fig:ppd_6x2} show histograms of the test statistic $T(D_2,\vartheta)$ computed from the data and from the realizations, $T(D_2^{\rm sim},\vartheta)$.  Fig.~\ref{fig:ppd_5x2} presents the distributions used to assess consistency between \3x2pt and $\galk+\sheartk$.  In this case, the data vector, $D_2$, is $\galk+\sheartk$, and we marginalize over the posterior from the \3x2pt analysis.  For  Fig.~\ref{fig:ppd_6x2}, the data vector is $\kk$ and we marginalize over the posterior from the \5x2pt analysis.  Note that the histograms of the test statistic have a finite extent because the PPD marginalizes over the posterior $P(\vartheta|D_1,M,I)$.  

\clearpage

\begin{figure*}
\includegraphics[width=0.95\textwidth]
{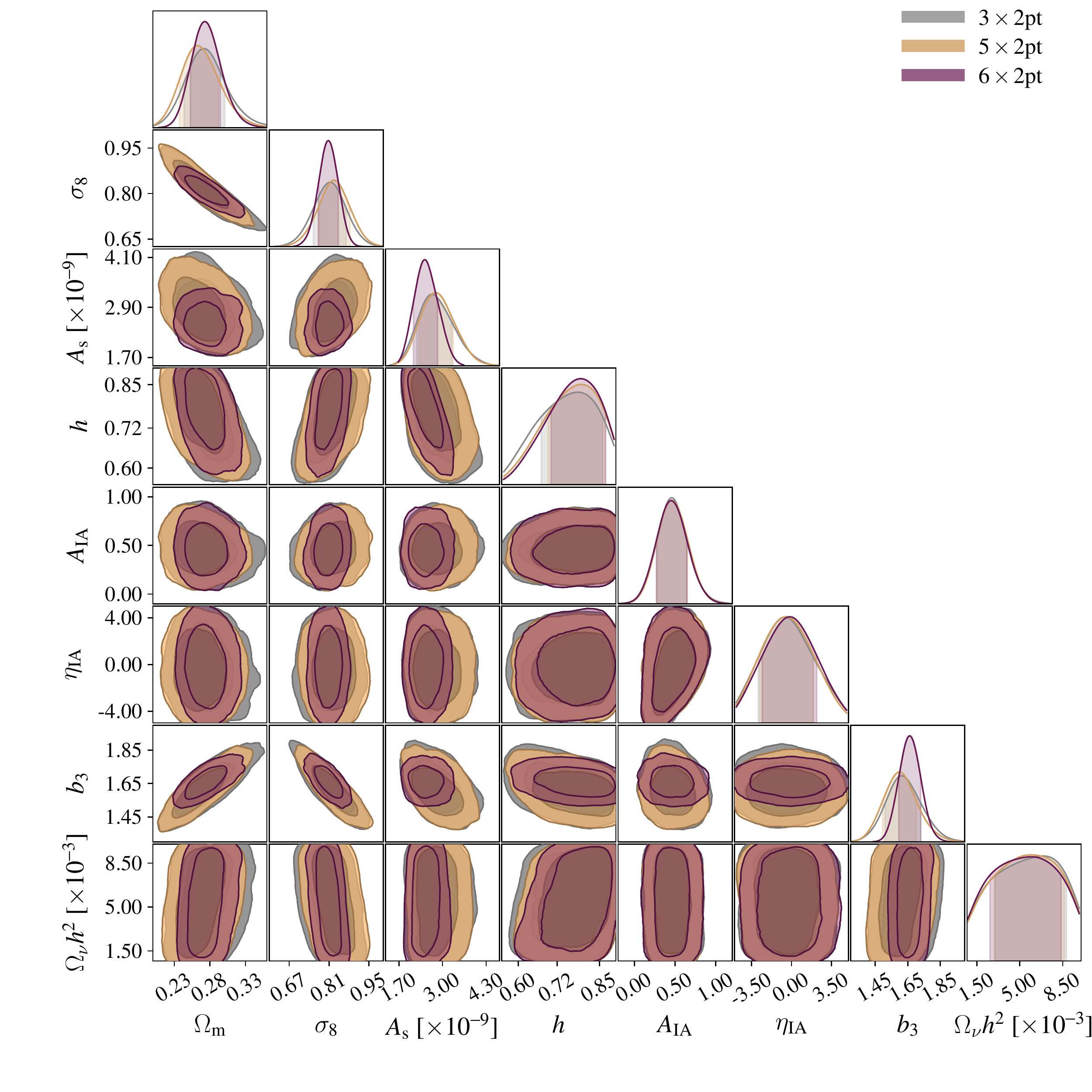}
\caption{Constraints on cosmological parameters ($\Omega_{\rm m}$, $\sigma_{8}$, $A_{\rm s}$, $h$), intrinsic alignment parameters ($A_{\rm IA}$, $\eta_{\rm IA}$) and galaxy bias for the third redshift bin ($b_{3}$). We choose to only show galaxy bias for one bin since all the bins show qualitatively the same behavior. The constraints are shown for different combinations of correlation functions: \3x2pt (gray), \5x2pt (gold) and \6x2pt (purple).}
\label{fig:more_params}
\end{figure*}

\section{Constraints on other parameters}
\label{sec:more_params}

In Fig.~\ref{fig:more_params}, we show the multi-dimensional parameter constraints from the three combinations of two-point correlation functions: \3x2pt, \5x2pt, and \6x2pt. Specifically, we show $A_{\rm s}$ (the amplitude of the matter power spectrum), $h$ (Hubble parameter), $A_{\rm IA}$ and $\eta_{\rm IA}$ (amplitude and redshift evolution of intrinsic alignment model), $b_{3}$ (one example of galaxy bias), and $\Omega_{\nu}$. We also examine the degeneracy of these parameters with our main cosmological constraints on $\Omega_{\rm m}$ and $\sigma_{8}$.

There is some improvement in constraints when going from \3x2pt to \5x2pt and a significant improvement going from \5x2pt to \6x2pt. For $h$ and the IA parameters, there is no sign of improvement going from \3x2pt to \5x2pt, and we do not expect any change going to \6x2pt since the CMB lensing auto-spectrum does not constrain IA (although in principle, degeneracy breaking could lead to some improvement). Improvements on constraints on galaxy bias has been discussed in Section \ref{sec:5x2constraints}.

\label{lastpage}

\end{document}